\documentclass[iop,apj,numberedappendix,appendixfloats,twocolappendix,revtex4]{emulateapj} 
\usepackage{times} 
\usepackage{latexsym,amsmath,amssymb,amsfonts} 
\usepackage{graphicx} 
\usepackage{fancyhdr}  
\usepackage{natbib} 
\usepackage{rotfloat} 
\usepackage{verbatim} 
\usepackage{rotating}   
\usepackage{url}
\usepackage{dcolumn}
\usepackage{morefloats} 
\usepackage[breaklinks,colorlinks,citecolor=blue,linkcolor=magenta]{hyperref}  
\begin{document}
\newcommand{\sqcm}{cm$^{-2}$}  
\newcommand{\lya}{Ly$\alpha$}
\newcommand{\lyb}{Ly$\beta$}
\newcommand{\lyg}{Ly$\gamma$}
\newcommand{\lyd}{Ly$\delta$}
\newcommand{\hi}{\mbox{\footnotesize H\,{\sc i}}}
\newcommand{\HI}{\mbox{H\,{\sc i}}}
\newcommand{\HII}{\mbox{H\,{\sc ii}}}  
\newcommand{\HH}{\mbox{H{$_2$}}}
\newcommand{\hh}{\mbox{\footnotesize H{$_2$}}} 
\newcommand{\HeI}{\mbox{He\,{\sc i}}}
\newcommand{\HeII}{\mbox{He\,{\sc ii}}}
\newcommand{\HeIII}{\mbox{He\,{\sc iii}}}  
\newcommand{\OI}{\mbox{O\,{\sc i}}}
\newcommand{\OII}{\mbox{O\,{\sc ii}}}
\newcommand{\OIII}{\mbox{O\,{\sc iii}}}
\newcommand{\OIV}{\mbox{O\,{\sc iv}}}
\newcommand{\OV}{\mbox{O\,{\sc v}}}
\newcommand{\OVI}{\mbox{O\,{\sc vi}}}
\newcommand{\OVII}{\mbox{O\,{\sc vii}}}
\newcommand{\OVIII}{\mbox{O\,{\sc viii}}} 
\newcommand{\CI}{\mbox{C\,{\sc i}}}
\newcommand{\CII}{\mbox{C\,{\sc ii}}}
\newcommand{\CIII}{\mbox{C\,{\sc iii}}}
\newcommand{\CIV}{\mbox{C\,{\sc iv}}}
\newcommand{\CV}{\mbox{C\,{\sc v}}}
\newcommand{\CVI}{\mbox{C\,{\sc vi}}}  
\newcommand{\SiII}{\mbox{Si\,{\sc ii}}}
\newcommand{\SiIII}{\mbox{Si\,{\sc iii}}}
\newcommand{\SiIV}{\mbox{Si\,{\sc iv}}}
\newcommand{\SiXII}{\mbox{Si\,{\sc xii}}}   
\newcommand{\SII}{\mbox{S\,{\sc ii}}}
\newcommand{\SIII}{\mbox{S\,{\sc iii}}}
\newcommand{\SIV}{\mbox{S\,{\sc iv}}}
\newcommand{\SV}{\mbox{S\,{\sc v}}}
\newcommand{\SVI}{\mbox{S\,{\sc vi}}}  
\newcommand{\NI}{\mbox{N\,{\sc i}}}   
\newcommand{\NII}{\mbox{N\,{\sc ii}}}   
\newcommand{\NIII}{\mbox{N\,{\sc iii}}}     
\newcommand{\NIV}{\mbox{N\,{\sc iv}}}   
\newcommand{\NV}{\mbox{N\,{\sc v}}}    
\newcommand{\PII}{\mbox{P\,{\sc ii}}} 
\newcommand{\PV}{\mbox{P\,{\sc v}}} 
\newcommand{\NeIV}{\mbox{Ne\,{\sc iv}}}   
\newcommand{\NeV}{\mbox{Ne\,{\sc v}}}   
\newcommand{\NeVI}{\mbox{Ne\,{\sc vi}}}   
\newcommand{\NeVII}{\mbox{Ne\,{\sc vii}}}   
\newcommand{\NeVIII}{\mbox{Ne\,{\sc viii}}}   
\newcommand{\NeIX}{\mbox{Ne\,{\sc ix}}}   
\newcommand{\NeX}{\mbox{Ne\,{\sc x}}} 
\newcommand{\MgI}{\mbox{Mg\,{\sc i}}}
\newcommand{\MgII}{\mbox{Mg\,{\sc ii}}}  
\newcommand{\MgX}{\mbox{Mg\,{\sc x}}}   
\newcommand{\AlII}{\mbox{Al\,{\sc ii}}}  
\newcommand{\FeII}{\mbox{Fe\,{\sc ii}}}  
\newcommand{\FeIII}{\mbox{Fe\,{\sc iii}}}   
\newcommand{\NaIX}{\mbox{Na\,{\sc ix}}}   
\newcommand{\ArVIII}{\mbox{Ar\,{\sc viii}}}   
\newcommand{\AlXI}{\mbox{Al\,{\sc xi}}}   
\newcommand{\CaII}{\mbox{Ca\,{\sc ii}}}  
\newcommand{\TiII}{\mbox{Ti\,{\sc ii}}}  
\newcommand{\MnII}{\mbox{Mn\,{\sc ii}}}  
\newcommand{\NaI}{\mbox{Na\,{\sc i}}}  
\newcommand{\ArI}{\mbox{Ar\,{\sc i}}}  
\newcommand{\zabs}{$z_{\rm abs}$}
\newcommand{\zmin}{$z_{\rm min}$}
\newcommand{\zmax}{$z_{\rm max}$}
\newcommand{\zqso}{$z_{\rm qso}$}
\newcommand{\zgal}{$z_{\rm gal}$}
\newcommand{\degree}{\ensuremath{^\circ}}
\newcommand{\lapp}{\mbox{\raisebox{-0.3em}{$\stackrel{\textstyle <}{\sim}$}}}
\newcommand{\gapp}{\mbox{\raisebox{-0.3em}{$\stackrel{\textstyle >}{\sim}$}}}
\newcommand{\be}{\begin{equation}}
\newcommand{\en}{\end{equation}}
\newcommand{\di}{\displaystyle}
\def\tworule{\noalign{\medskip\hrule\smallskip\hrule\medskip}} 
\def\onerule{\noalign{\medskip\hrule\medskip}} 
\def\bl{\par\vskip 12pt\noindent}
\def\bll{\par\vskip 24pt\noindent}
\def\blll{\par\vskip 36pt\noindent}
\def\rot{\mathop{\rm rot}\nolimits}
\def\alf{$\alpha$}
\def\refff{\leftskip20pt\parindent-20pt\parskip4pt}
\def\kms{km~s$^{-1}$}
\def\zem{$z_{\rm em}$}
\shorttitle{\HH\ absorption from a galaxy halo}        
\shortauthors{S. Muzahid et al.} 
\title{Molecular hydrogen absorption from the halo of a $z\sim$0.4 galaxy}          
\author{Sowgat Muzahid\altaffilmark{1},  
Glenn G. Kacprzak\altaffilmark{2},  
Jane C. Charlton\altaffilmark{1}, and  
Christopher W. Churchill\altaffilmark{3}  
}    
\affil{\\ $^{1}$The Pennsylvania State University, State College, PA 16802, USA; (sowgatm@gmail.com) \\  
$^{2}$Swinburne University of Technology, Victoria 3122, Australia \\   
$^{3}$New Mexico State University, Las Cruces, NM 88003, USA \\ 
}

\begin{abstract}  

Lyman- and Werner-band absorption of molecular hydrogen (\HH) is detected in $\sim$50\% of low-redshift ($z<1$) DLAs$/$sub-DLAs with $N(\HH)>10^{14.4}$~\sqcm. However, the true origin(s) of the \HH-bearing gas remain elusive. Here we report a new detection of an \HH\ absorber at \zabs~$=$~0.4298 in the $HST/$COS spectra of quasar PKS~2128--123. The total $N(\HI)$ of $10^{19.50\pm0.15}$~\sqcm\ classifies the absorber as a sub-DLA. \HH\ absorption is detected up to the $J=3$ rotational level with a total $\log N(\HH)=$~16.36$\pm$0.08 corresponding to a molecular fraction of $\log f_{\hh}=$~$-2.84\pm0.17$. The excitation temperature of $T_{\rm ex}=$~206$\pm$6~K indicates the presence of cold gas. Using detailed ionization modeling we obtain a near-solar metallicity (i.e., $\rm [O/H]=$$-0.26\pm0.19$) and a dust-to-gas ratio of $\log \kappa\sim-0.45$ for the \HH-absorbing gas. The host galaxy of the sub-DLA is detected at an impact parameter of $\rho\sim$48~kpc with an inclination angle of $i\sim$48\degree\ and an azimuthal angle of $\Phi\sim$15\degree\ with respect to the QSO sightline. We show that co-rotating gas in an extended disk cannot explain the observed kinematics of \MgII\ absorption. Moreover, the inferred high metallicity is not consistent with the scenario of gas accretion. An outflow from the central region of the host galaxy, on the other hand, would require a large opening angle (i.e., 2$\theta>$150\degree), much larger than the observed outflow opening angles in Seyfert galaxies, in order to intercept the QSO sightline. We thus favor a scenario in which the \HH-bearing gas is stemming from a dwarf-satellite galaxy, presumably via tidal and$/$or ram pressure stripping. Detection of a dwarf galaxy candidate in the $HST/$WFPC2 image at an impact parameter of $\sim$12~kpc reinforces such an idea.  
\end{abstract}  
\maketitle
\section{Introduction} 
\label{sec:intro}  

Molecular hydrogen (\HH) is the most abundant molecule in the universe, and it plays a crucial role in star formation in the interstellar medium \citep[ISM; see][for a review]{Shull82}. \HH\ exhibits numerous transitions from its ground electronic state to the Lyman and Werner bands at ultraviolet (UV) wavelengths (900--1130~\AA). The Lyman- and Werner-band absorption of \HH\ is detected in diverse astronomical environments, e.g., in the Galactic disk \citep[]{Spitzer75, Savage77}, Galactic halo \citep[]{Gillmon06,Wakker06}, Magellanic Clouds \citep[]{Tumlinson02, Welty12}, Magellanic Stream \citep[MS;][]{Sembach01}, Magellanic Bridge \citep[MB;][]{Lehner02}, high-velocity clouds \citep[HVCs;][]{Richter99b,Richter01a}, intermediate-velocity clouds \citep[IVCs;][]{Gringel00,Richter03}, and in high-redshift damped \lya\ absorbers \citep[DLAs;][]{Ledoux03,Noterdaeme08}.  

For high-redshift ($z>1.8$) absorbers, the UV transitions of \HH\ get redshifted to optical wavelengths. Absorption lines of high-$z$ \HH\ absorbers, thus, are well studied using high-resolution optical spectra obtained from ground-based telescopes \citep[e.g.,][]{Srianand05,Srianand08,Srianand12}. Due to the lack of high-resolution, UV-sensitive, space-based spectrograph, \HH\ was never studied in absorption at low-$z$ ($z<1$) beyond the Local Group until the last decade. However, the high throughput of the Cosmic Origins Spectrograph (COS) on board the {\it Hubble Space Telescope} ($HST$) has recently enabled such observation. Detection and analysis of low-$z$ \HH\ absorbers have been presented in several recent studies \citep[i.e.,][]{Crighton13,Oliveira14,Srianand14,Dutta15,Muzahid15a}. 

Using archival $HST/$COS spectra, \cite{Muzahid15a} have surveyed \HH\ absorption in 27 low-$z$ DLAs$/$sub-DLAs for the first time and have reported detections of 10 \HH\ absorbers in total. The \HH\ detection rate at low-$z$, i.e. $50^{+25}_{-12}$\%, for systems with $N(\HH)>10^{14.4}$~\sqcm, is a factor of $\gtrsim2$ higher than that found at high-$z$ by \citet{Noterdaeme08}. The increase of the cosmic mean metallicity of DLAs$/$sub-DLAs \citep[e.g.,][]{Rafelski12,Som13} and the dimming of the ambient radiation field due to the decrease of the cosmic star formation rate density with cosmic time \citep[e.g.,][]{Bouwens11,Haardt12} are thought to be responsible for such an enhanced \HH\ detection rate at low-$z$. 

The median $N(\HI)$ value of $10^{19.5}$~\sqcm\ for the low-$z$ sample of \cite{Muzahid15a} is an order of magnitude lower than that of the high-$z$ sample. Nevertheless, the low-$z$ \HH\ absorbers show molecular fractions, \linebreak $f_{\hh} = 2N(\HH)/[N(\HI)+2N(\HH)]$, that are comparable to the high-$z$ \HH\ absorbers (e.g., median $\log f_{\hh} = -1.93\pm0.63$). High molecular fractions at lower $N(\HI)$ values indicate that (a) the density is extremely high and$/$or (b) the prevailing radiation field is very weak in the absorbing gas.    

Using simple photoionization models and from the lack of high $J$ (i.e., $J>3$) excitations, \cite{Muzahid15a} have inferred that the radiation field is much weaker than that seen in the Milky Way diffuse ISM \citep[]{Black87}, in contrast to high-$z$ \HH\ absorbers \citep[e.g.,][]{Srianand05}. The density of the \HH-absorbing gas is predicted to be in the range of 10--100~cm$^{-3}$ (or lower), which is consistent with that of the Galactic diffuse ISM. The median rotational excitation temperature of $T_{01} = 133\pm55$~K\footnote{Defined as $\frac{N_1}{N_0} = \frac{g_1}{g_0}~{\rm exp}(-E_{01}/kT_{01})$, where $N_i$ and $g_i$ are the column density and statistical weight of the $i$th rotational level, respectively.} is also suggestive of a diffuse atomic gas-phase \citep[]{Snow06}.  

Detecting host galaxies of high-$z$ \HH\ absorbers is difficult, and searches have not been successful so far. The advantage of studying low-$z$ \HH\ absorbers is that it is relatively easier to identify the host galaxies responsible for absorption. For example, 8 of the 10 \HH\ absorbers in the sample of \cite{Muzahid15a} have host galaxies identified from the literature. Interestingly, 5 of the 8 \HH\ absorbers have host galaxies at an impact parameter of $\rho>20$~kpc. Moreover, the two systems with the highest molecular fractions show the largest impact parameters \citep[i.e. $\rho>50$~kpc; see Fig.~9 of][]{Muzahid15a}. Even for the lowest impact parameter ($\rho <10$) system the QSO sightline is well beyond the stellar disk of the host galaxy \citep[see, e.g.,][]{Petitjean96,Chen05}. It is, therefore, conjectured that the low-$z$ \HH\ absorbers are not related to star-forming disks, but emerge from tidally stripped or ejected disk material in the halo.   

Here we present a detailed analysis of a new \HH\ absorber in a sub-DLA at \zabs~$=$~0.4298 detected toward QSO~PKS~2128--123. The host galaxy of the \HH\ absorber is detected at $\rho\sim$48~kpc. The main focus of this work is to understand the origin of the detected \HH\ absorption. The article is organized as follows: In Section~\ref{sec:obs} we present the available observations. Analysis related to absorption spectra and ionization models are presented in Section~\ref{sec:absana}. In Section~\ref{sec:galana} analysis of the host galaxy is presented. Our results are discussed in Section~\ref{sec:diss} followed by a summary in Section~\ref{sec:summ}. Throughout this paper we adopt an $H_{0}=70$~\kms~Mpc$^{-1}$, $\Omega_{\rm M}=0.3$, and $\Omega_{\rm \Lambda}=0.7$ cosmology. Solar abundances of heavy elements are taken from \citet{Asplund09}. All the distances given are proper (physical) distances.


\begin{figure*} 
\centerline{
\vbox{
\centerline{\hbox{ 
\includegraphics[width=0.95\textwidth]{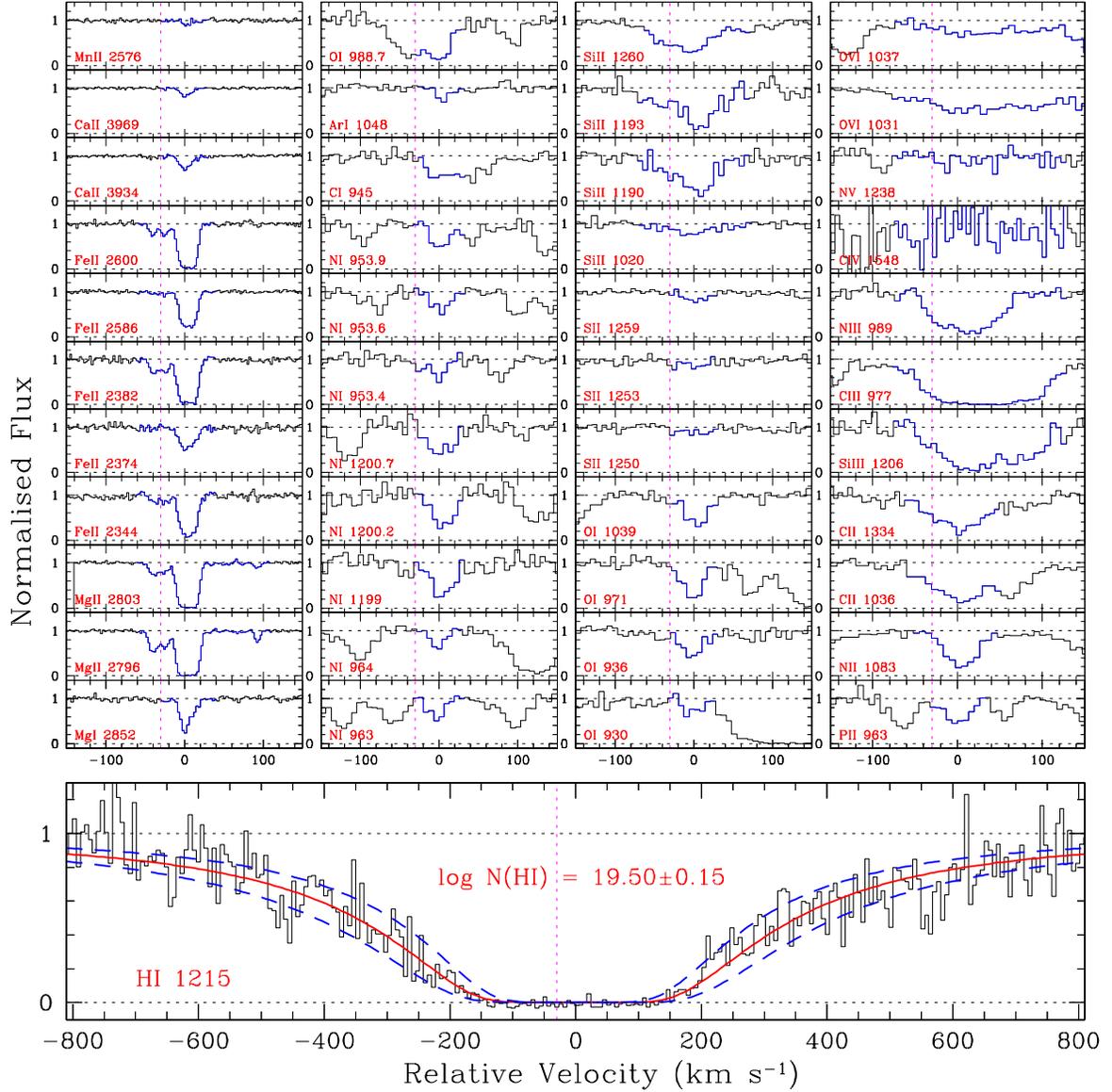}   
}} 
}} 
\vskip-0.9cm  
\caption{Velocity plot of the sub-DLA system studied here. The zero velocity corresponds to the redshift of the strongest metal-line component, i.e., \zabs\ $=$~0.429805. The vertical dotted line in each panel corresponds to the host galaxy redshift of \zgal~$=$~0.42966. Bottom: Sub-damped \lya\ profile of the absorber. The black histograms represent the data. The smooth (red) and dashed (blue) curves represent the best-fitting Voigt profile and its 1$\sigma$ uncertainty, respectively. Top: Various metal lines arising from the same system. The absorption lines plotted in the leftmost panel are detected in the HIRES spectra. All other absorption lines are from COS spectra. The relevant part of each line is marked in blue in order to distinguish it from an unrelated absorption. Note that \NV\ and \CIV\ are not detected in this system.  
}  
\label{fig:vplot}   
\end{figure*} 

\vskip1cm  
\section{Observations}    
\label{sec:obs}  

\subsection{Absorption Data}  
\subsubsection{$HST/$COS} 

Far-ultraviolet (FUV) spectra of the background QSO PKS~2128--123 were obtained using $HST/$COS Cycle-21 observations under program ID GO-13398 as a part of our ``Multiphase Galaxy Halos" survey. The properties of COS and its in-flight operations are discussed by \cite{Osterman11} and \cite{Green12}. The observations consist of G130M and G160M FUV grating integrations at a medium resolution of $R\sim$20,000 (FWHM$\sim$18~\kms) over the wavelength range of 1140--1800~\AA. However, due to the complete Lyman-limit break from the absorber of interest  at \zabs~$=$~0.4298, no QSO flux is recorded for $\lambda <$~1310~\AA. The data were retrieved from the $HST$ archive and reduced using the STScI {\sc calcos} v2.21 pipeline software. The reduced data were flux calibrated. To increase the spectral signal-to-noise ratio (S$/$N), individual exposures were aligned and co-added using the IDL code ``{\it coadd\_x1d}" developed by \cite{Danforth10}\footnote{http://casa.colorado.edu/$\sim$danforth/science/cos/costools.html}. The combined spectrum has a S$/$N~$\sim$9--20 per resolution element. As the COS FUV spectra are significantly oversampled (i.e. 6 raw pixels per resolution element), we binned the data by 3 pixels. This improves the S$/$N per pixel by a factor of $\sqrt3$. All our measurements and analyses were performed on the binned data. 

In addition to the FUV observations, near-UV (NUV) spectral data are also available for the QSO. These NUV observations consist of G185M and G225M grating integrations. The G185M grating data were obtained using $HST/$COS Cycle-19 observations under program ID GO-12536, whereas the G225M grating data were obtained by our own program, ID GO-13398. Individual G185M and G225M integrations were co-added using custom-written IDL code following similar algorithms to the ``{\it coadd\_x1d}". Both G185M and G225 data consist of three 35~\AA\ stripes separated by two 65~\AA\ gaps. G185M data cover 1670--1705~\AA, 1770--1805~\AA, and 1870--1905~\AA, whereas G225M data cover 2100--2135~\AA, 2200--2235~\AA, and 2300--2335~\AA. The co-added G185M spectrum has an S$/$N of 5--13 per resolution element. However, the G225M spectrum has an S$/$N~$<3$ per resolution element.

\subsubsection{Keck$/$HIRES} 

The optical spectrum of PKS~2128--123 was obtained with the HIRES mounted on the Keck-I 10~m telecope on Mauna Kea in Hawaii. The QSO was observed under program IDs C54H (PI: W. Sargent), U51H (PI: S. Vogt), and C99H (C. Steidel). We use the spectrum obtained from the Keck Observatory Archive's automated reduction pipeline software for our analysis. The pipeline-reduced spectrum was not flux calibrated; however, the wavelength was both vacuum and heliocentric velocity corrected. The final reduced spectrum covered 3200--6100~\AA\ at a spectral resolution of $R\sim$45,000 (FWHM~$\sim$6.6~\kms) with a spectral S$/$N of $\sim$20 and $\sim$45 per pixel in the blue and red parts, respectively. We note that the pipeline-reduced spectrum used here is identical to the one independently reduced and kindly provided to us by Hadi Rahmani using {\sc makee}\footnote{http://www.astro.caltech.edu/$\sim$tb/makee/About.html} data reduction software. Continuum normalizations of all the spectra (COS$/$FUV, NUV and Keck$/$HIRES) were done by fitting the line-free regions with smooth low-order polynomials.

\subsection{Galaxy Data} 
\label{sec:galdata}   

A 600 s $HST/$WFPC2 F702W image of the PKS~2128--123 field was obtained under program ID 5143. The reduced and calibrated image was obtained from the WFPC2 Associations Science Products Pipeline. Apparent Vega magnitudes were determined from 1.5$\sigma$ isophotes using SExtractor \citep{Bertin96}. Galaxy sky orientation parameters, such as inclination angle ($i$) and azimuthal angle ($\Phi$), were determined using GIM2D models \citep{Simard02} following the methods of \citet{Kacprzak11}. The image has a limiting magnitude of 25.5, which translates to a $B$-band absolute magnitude of $M_{B} = -15.2$ and $L = 0.004 L_{\ast}$ at $z = 0.43$.    

The optical spectrum of the host galaxy was obtained using the Keck Echelle Spectrograph and Imager \citep[ESI;][]{Sheinis02} on 2015 July 16 with an exposure time of 6000 s. The mean seeing was $0.9''$ (${\rm FWHM}$) with variable cloud cover. The ESI slit is $20''$ long and $1''$ wide and we used $2\times2$ on-chip CCD binning. Binning by two in the spatial direction results in pixel sizes of $0.27''-0.34''$ over the echelle orders of interest. Binning by two in the spectral direction results in a velocity dispersion of $22$~\kms~pixel$^{-1}$ (${\rm FWHM}\sim90$~\kms).    

The spectrum was reduced using the standard echelle package in {\sc iraf}\footnote{{\sc iraf} is distributed by the National Optical Astronomy Observatory, which is operated by the Association of Universities for Research in Astronomy (AURA) under a cooperative agreement with the National Science Foundation.} along with standard calibrations and was vacuum and heliocentric velocity corrected. The data were not flux calibrated due to the variable conditions of the sky. We use our own fitting software {\sc fitter} \citep[see][]{Churchill00a}, which computes best-fit Gaussian amplitudes, line centers, and widths to obtain emission-line redshifts and equivalent widths.     

The galaxy rotation curve was extracted following the methods of \citet{Kacprzak10a} where we extract individual spectra by summing 3-pixel-wide apertures at 1-pixel spatial increments along the slit. Each spatial spectrum is then wavelength calibrated by extracting spectra of the arc line lamps at the same spatial pixels as the extracted galaxy spectra. The centroid of each emission line in each spectrum was determined with a Gaussian fit using {\sc fitter}.

\begin{figure*} 
\centerline{
\vbox{
\centerline{\hbox{ 
\includegraphics[width=0.90\textwidth]{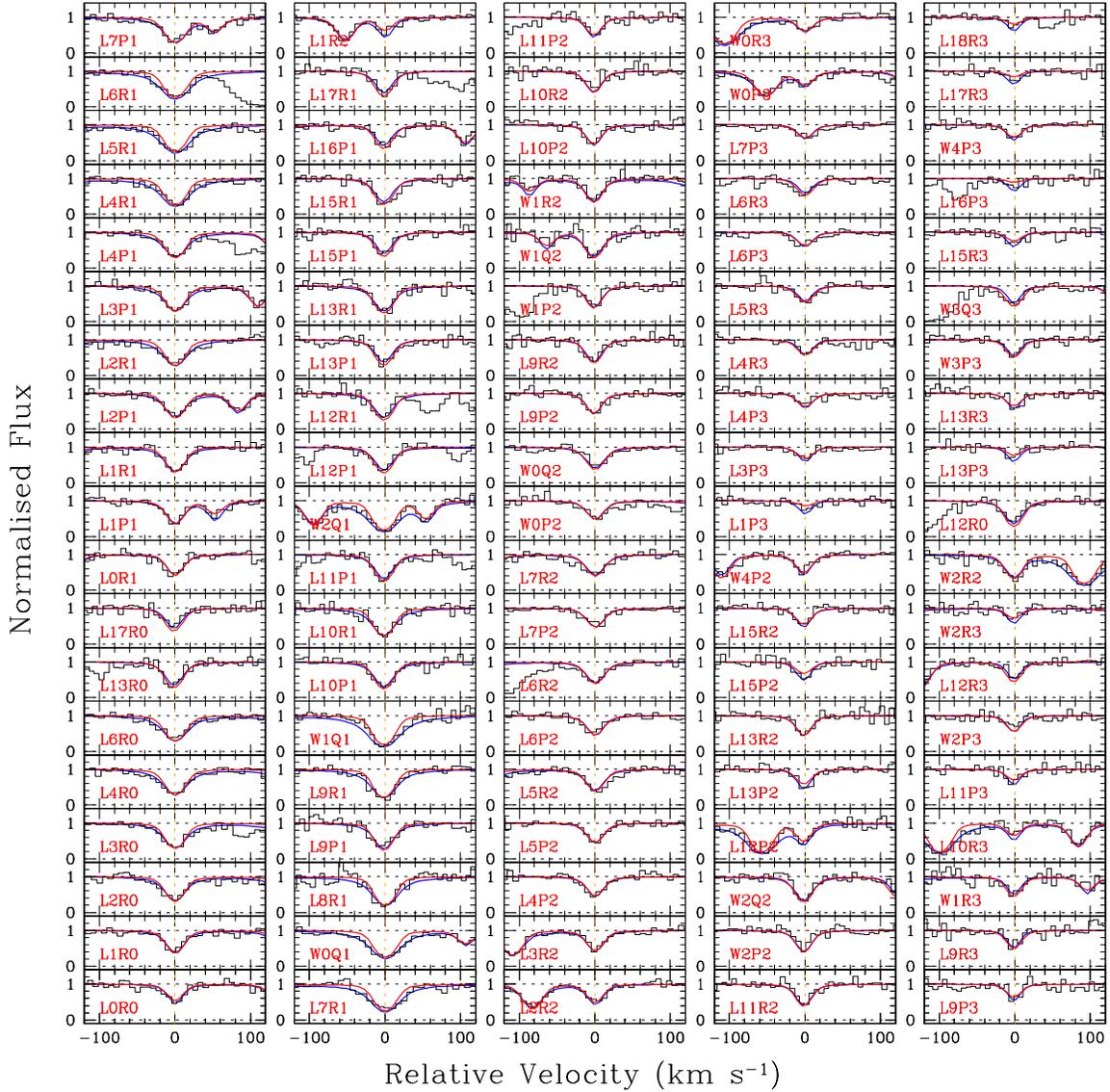}   
}} 
}}  
\vskip-0.3cm  
\caption{Velocity plot of numerous \HH\ absorption lines from different $J$ levels arising from the sub-DLA. The zero velocity corresponds to \zabs~$=$~0.429805. The red and blue smooth curves are the best-fitting Voigt profiles to the data (black histograms) corresponding to the two different {\sc vpfit} solutions as discussed in the text. The blue curves correspond to a low $b$-parameter (2.8$\pm$0.3~\kms) solution. The red curves represent a solution with $b(\HH)=$~7.1$\pm$0.3~\kms. The low $b$-parameter solution results in an $N(\HH)$ of $\sim$2 orders of magnitude higher as compared to the other. A $b$-parameter of 2.8~\kms\ is significantly lower than the spectral resolution of COS. Moreover, such a low $b$-value is not observed for the metal lines detected at the same velocity in the high-resolution Keck/HIRES data. We therefore prefer the solution with $b=$~7.1~\kms, i.e., the red curves. Our photionization model (see Section~\ref{sec:mod}) further suggests that the $N(\HH)$ corresponding to the low $b$-parameter requires a density inconsistent with the observed $N(\MgI)/N(\MgII)$ ratio.                      
}      
\label{fig:H2fit}     
\end{figure*} 

\begin{figure*} 
\centerline{
\vbox{
\centerline{\hbox{ 
\includegraphics[width=0.85\textwidth]{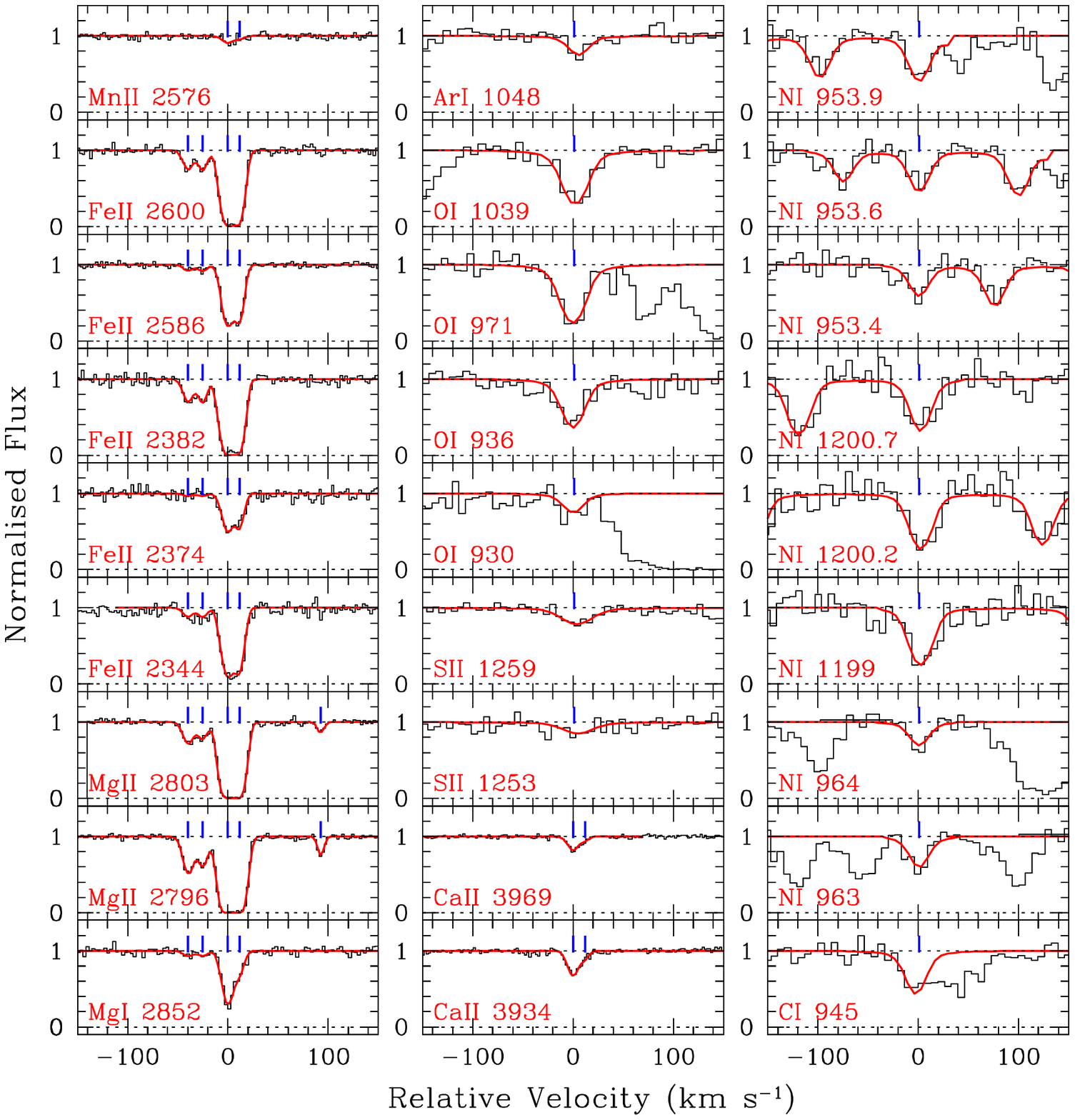}  
}} 
}}  
\vskip-0.9cm  
\caption{Voigt profile fits to the metal lines arising from the sub-DLA. The smooth red curves are the model profiles overplotted on top of data (black histograms). The zero velocity corresponds to \zabs~$=$~0.429805. The line centroids of individual components used to fit a given line are marked by the blue ticks.}        
\label{fig:fitHIRES}   
\end{figure*} 


\section{Absorption Analysis} 
\label{sec:absana}

\subsection{Description of the Absorber}  
\label{subsec:des}  

In Figure~\ref{fig:vplot} we display the absorption profiles of \lya\ and different metal lines arising from the \zabs~$=$~0.4298 absorber. The \lya\ absorption clearly shows a damping wing. Strong, saturated absorption lines from \MgII$\lambda\lambda$2796,2803 doublets are detected in the high-resolution Keck spectrum. The very weak component at $v\sim+$90~\kms\ is not detected in any other neutral$/$singly ionized metal transitions. The two other weak components at $v\sim-$40 and $-$25 \kms\ are, however, present in the \MgI\ and \FeII\ absorption lines. The new COS observations cover several metal lines stemming from different elements (e.g., C, N, O, S, Si) and from different ionization states (e.g., \OI, \CIII, \OVI). Except for the \CI\footnote{The red wing of the \CI$\lambda$945 line is partially blended, possibly with the \lya\ absorption from the \zabs$=$0.1118 system.}, all other unblended neutral species (i.e., \OI\footnote{The \OI$\lambda$988.7 line is self-blended with the weaker \OI$\lambda$988.6 and $\lambda$988.5 lines}, \NI, and \ArI) exhibit a single-component, simple absorption kinematics. Except for the \SiII$\lambda$1020, \PII$\lambda$963, and \NII$\lambda$1083\footnote{We note that the \NII\ line is partially blended with the Galactic \CIV$\lambda$1550 line.}, presence of multicomponent structure is seen in all other singly ionized and higher-ionization absorption lines\footnote{Note that the red wing of the \CII$\lambda$1036 is contaminated with the L5R0 \HH\ line detected from this sub-DLA. Moreover, the \NIII$\lambda$989 line is partly blebded with the \SiII$\lambda$989.8 absorption.}. Very strong, saturated \CIII\ absorption is detected over the entire velocity range for which other low-ionization species, \MgII\ in particular, are detected.  

The details of the high-ionization species are presented in Appendix~\ref{sec:highion} since the main focus of this paper is the low-ionization gas-phase giving rise to \HH\ absorption. In brief, \OVI$\lambda\lambda$1031,1037 absorption lines are detected in both the doublets in the velocity range of $-100$ to $+200$~\kms\ with a total line spread of $\Delta v_{90}=$~246 \kms\ (see Figure~\ref{fig:highion}). No \CIV\ and$/$or \NV\ lines are detected in this system. The Voigt profile fit parameters for the \OVI\ absorption and 3$\sigma$ upper limits on $N(\NV)$ and $N(\CIV)$ are given in Table~\ref{tab:highion}.      

Besides several metal absorption lines, numerous Lyman- and Werner-band absorption lines of \HH, arising from the $J=$ 0, 1, 2, and 3 rotational levels, are detected from this absorber. The \HH\ absorption lines that are not self-blended or contaminated by other unrelated absorption are shown in Figure~\ref{fig:H2fit}. Similar to neutral metal-line transitions detected in the COS spectra, the \HH\ lines from different $J$ levels also show a single-component, simple absorption kinematics. Importantly, the absorption redshifts of \HH\ transitions coincide with the strongest metal-line component.                          

\subsection{Absorption Line Measurements}    
\label{subsec:mea}  

We primarily use the {\sc vpfit}\footnote{http://www.ast.cam.ac.uk/$\sim$rfc/vpfit.html} software for measuring absorption line parameters (i.e., $N$, $b$, and $z$). Additionally, whenever possible, we have used the curve-of-growth (COG) and the apparent optical depth (AOD) techniques. First, we fit all the metal lines detected in the high-resolution Keck spectrum simultaneously, i.e., by keeping the absorption redshift and $b$-parameter of a given component tied to each other. A minimum of five components are required to fit the \MgII\ doublets satisfactorily. As noted earlier, the weak component at $v\sim+$90~\kms\ is not present in any other lines detected in the Keck spectrum. Except for the component at $0$~\kms, we tied the $b$-parameter of different ions via thermal broadening with a gas temperature of $10^{4}$~K. We note that a gas cloud photoionized by the extragalactic UV background radiation cannot cool efficiently below $10^{4}$~K via metal-line cooling \citep[]{Wiersma09}. Therefore, it is expected that the components without detected \HH\ will have gas temperatures on the order of or higher than $10^{4}$~K. For the strongest metal-line component at $0$~\kms, we set the temperature to be 156~K, i.e., the rotational excitation temperature ($T_{01}$) we obtain from the \HH\ column densities in the $J=0$ and $J=1$ levels as discussed below. The Voigt profile fit parameters are summarized in Table~\ref{tab:fitHIRES}, and the model profiles are shown in Figure~\ref{fig:fitHIRES}. Since there is no \lya\ forest crowding at low-$z$ and the absorption lines of interest do not fall on top of any emission line, the continuum fitting uncertainty is generally small and is not taken into account in the absorption-line fit parameters. 

Next, we fit the sub-DLA profile detected in the medium-resolution COS spectrum. Note that the line-spread function (LSF) of the COS is not a Gaussian. For our Voigt profile fit analysis we adopt the latest LSF given by \citet{Kriss11}. The LSF was obtained by interpolating the LSF tables at the observed central wavelength for each absorption line and was convolved with the model Voigt profile while fitting absorption lines using {\sc vpfit}. The total \HI\ column density we obtain by fitting the damping wing (Figure~\ref{fig:vplot}) of the \lya\ absorption is $N(\HI) = 10^{19.50\pm0.15}$~\sqcm.

Following \cite{Muzahid15a}, we choose a set of uncontaminated \HH\ lines for Voigt profile fitting (see Figure~\ref{fig:H2fit}). We fit them simultaneously by keeping the $b$-parameter tied. Additionally, we constrain the column density to be the same for all the transitions from a given $J$ level and allow it to be different for different $J$ levels. The COS wavelength calibration is known to have uncertainties at the level of 10--15~\kms\ \citep[]{Savage11a}. We use the numerous \HH\ absorption lines as guides to improve the wavelength calibration uncertainty. Our final corrected spectrum has a velocity accuracy of $\sim\pm$5~\kms. As a consequence, we did allow the redshift of individual transitions to be different from each other by a maximum of $\pm$5~\kms.

We noticed that, depending on the initial-guess $b$-parameter, {\sc vpfit} converges at two different solutions with significantly different total $N(\HH)$ values. The fit parameters are summarized in Table~\ref{tab:H2fit}. For an initial-guess $b$-value of $\ge$8~\kms, {\sc vpfit} converges with a $b(\HH)$ of 7.1$\pm$0.3 and a total $\log N(\HH)=$~16.36$\pm$0.08. On the other hand, an initial-guess $b$-value of $<$8~\kms\ leads to a solution with a $b(\HH)$ of 2.8$\pm$0.3 and a total $\log N(\HH)=$~18.27$\pm$0.03. Such a degeneracy is also present when we use the standard COG technique. The model profiles corresponding to the two different solutions are shown in Figure~\ref{fig:H2fit}. From the model profiles and$/$or from the reduced $\chi^2$ values, it is difficult to choose one model as opposed to the other. However, we note that the $b$-parameter in the latter case is significantly lower as compared to the velocity resolution of the COS spectra ($\sim$18~\kms). Such a narrow $b$-value is not measured in the corresponding metal-line component, even in the high-resolution Keck data. Furthermore, we demonstrate in the next section that the observed $N(\MgI)$ to $N(\MgII)$ ratio does not allow $N(\HH)$ to be as high as $10^{18.27}$~\sqcm. As such, we exclude the higher $N(\HH)$ solution.    

The unblended lines of \NI, \OI, \ArI, and \SII, detected in the COS spectra, are fitted with a single-component Voigt profile keeping both redshift and $b$-parameter free (see Table~\ref{tab:fitCOS}). The weak \ArI\ absorption is detected only in the $\lambda$1048 transition\footnote{The expected wavelength range for the other weaker transition of \ArI\ (i.e., $\lambda$1066) is noisy and partly blended with the $\rm L3P2$ line of \HH.}. A free fit to the \ArI$\lambda$1048 leads to a highly erroneous solution. We, therefore, fix the $b$-parameter at 6.3~\kms, which we obtained from the fit to the strongest metal-line component detected in the Keck spectrum. The $b$-parameter of 16.8$\pm$4.0 \kms\ required for a free fit to the weak \SII\ absorption lines is higher than that is observed for the other lines. Nevertheless, the estimated column density is robust since the lines are weak and fall on the linear part of COG. Finally, due to the blend in the red wing of the \CI\ absorption, we estimate the maximum column density that can be accommodated by keeping the $b$-parameter and component velocity (i.e., \zabs) the same as the \MgI\ line. We treat the $N(\CI)$ obtained by this method as a conservative upper limit.   

\begin{figure*} 
\centerline{
\vbox{
\centerline{\hbox{ 
\includegraphics[width=0.25\textwidth]{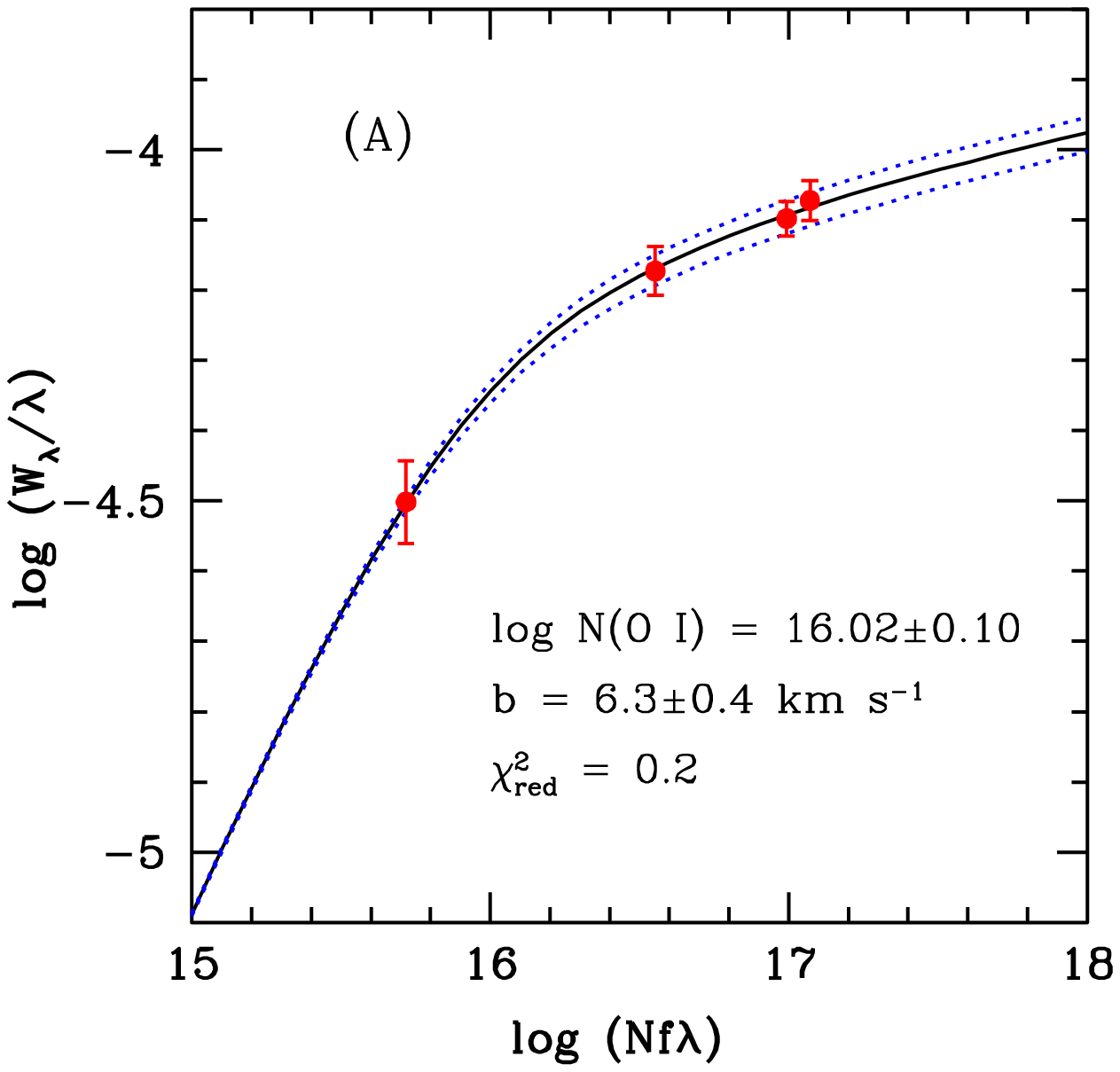} 
\hskip-0.2cm  
\includegraphics[width=0.25\textwidth]{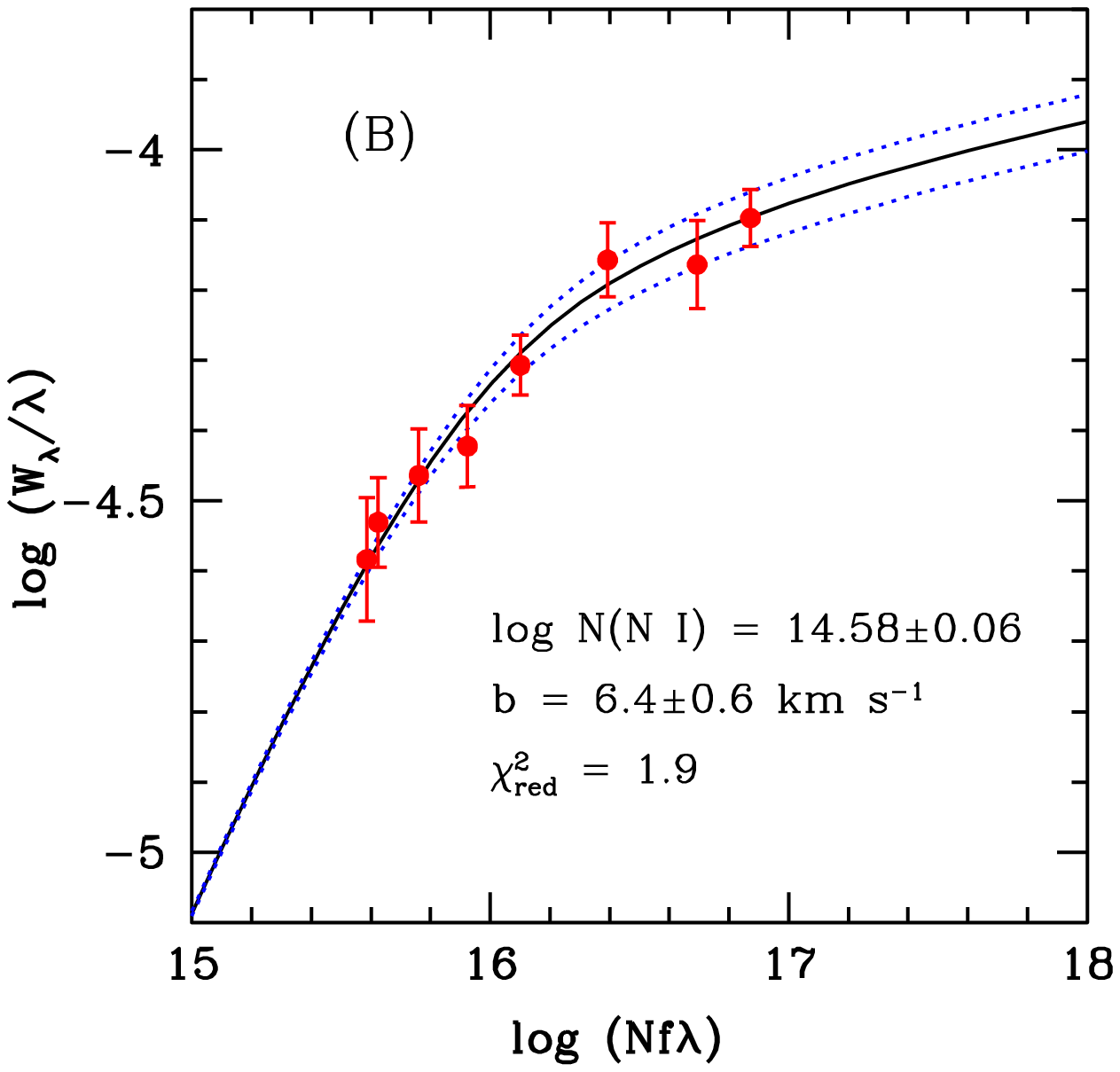} 
\hskip-0.2cm  
\includegraphics[width=0.25\textwidth]{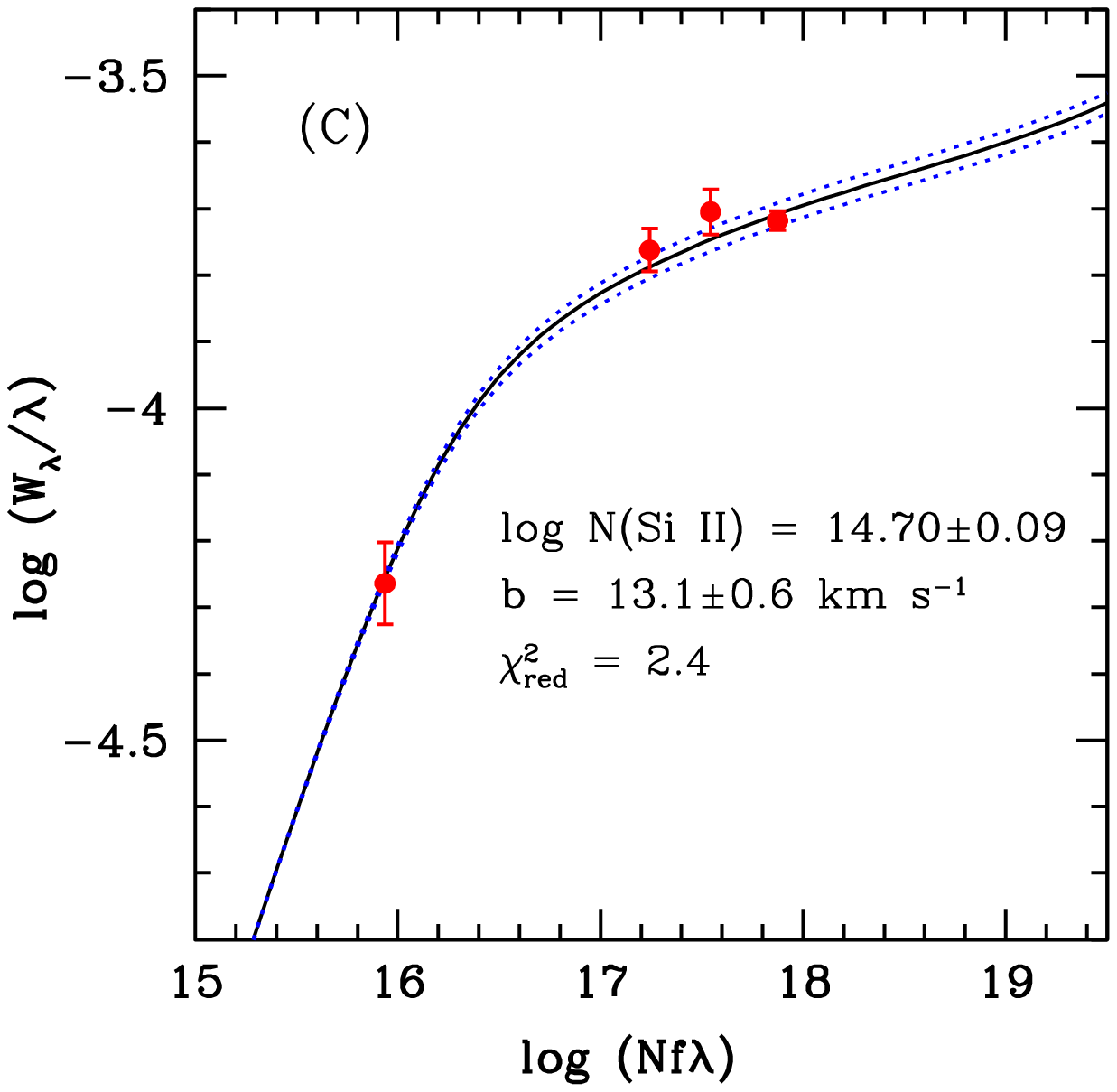}    
\hskip-0.2cm  
\includegraphics[width=0.25\textwidth]{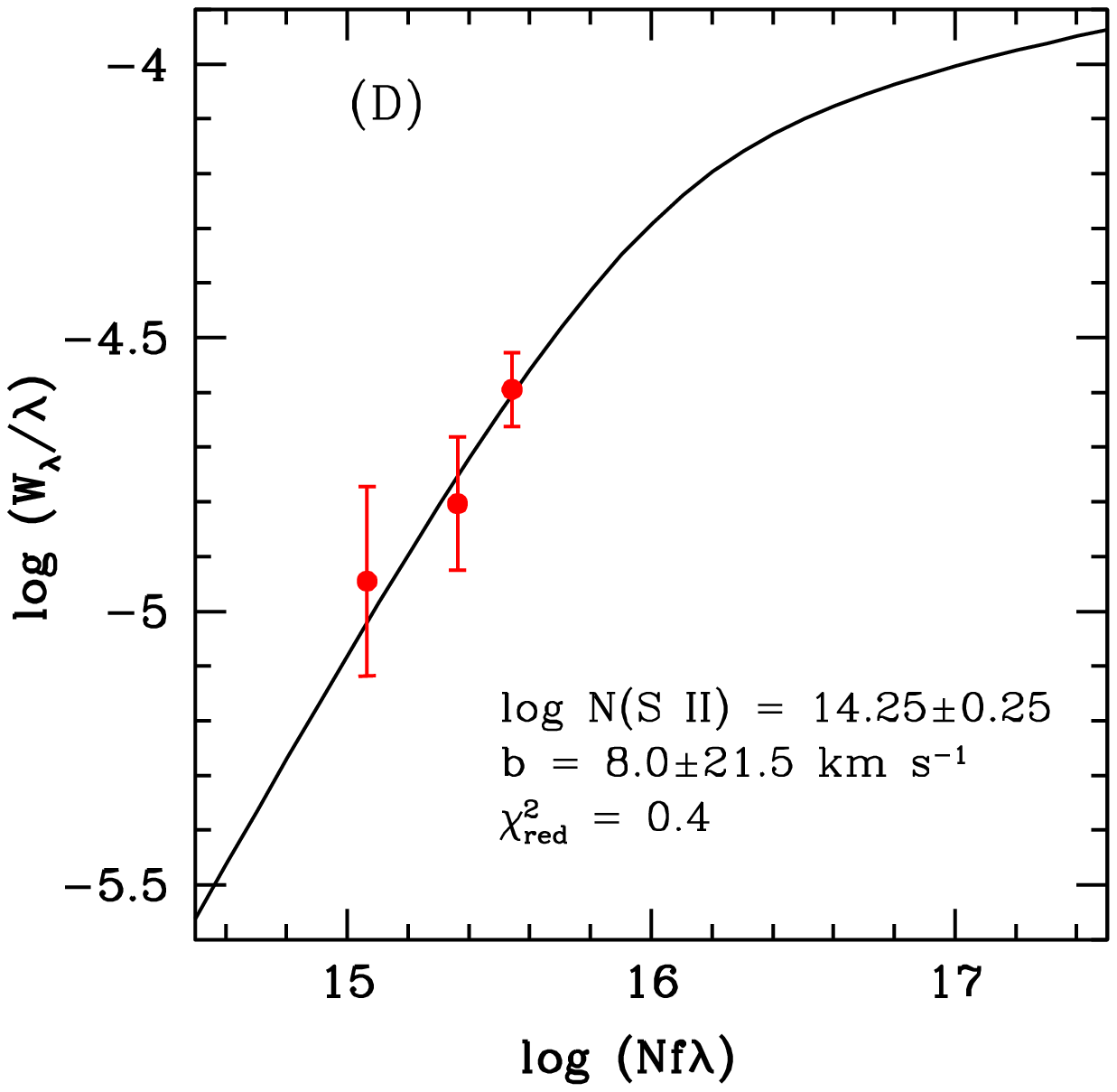}  
}} 
}}  
\caption{Results of COG analysis for \OI, \NI, \SiII, and \SII\ in panels A, B, C, and D, respectively. 
Note that for each case at least one transition is on the linear part of the COG, which enables us to 
estimate the total column density robustly. This is further supported by the independent AOD 
measurements of the weak lines (see Table~\ref{tab:summaryN}) that are on the linear part of the COG.}       
\label{fig:COG}  
\end{figure*}

Next, we perform COG analysis for the ions detected in the medium-resolution COS spectra with at least three available unblended transitions. The results of our COG analyses are shown in Figure~\ref{fig:COG}. The best-fitting COG column densities and $b$-parameters are given in each panel of the figure. The $b$-parameters obtained for both \OI\ and \NI, using the COG method, are consistent with the $b$-parameter we measure for the strongest absorption component in the Keck spectrum. Moreover, they are also consistent with the values we obtain using {\sc vpfit} within 1$\sigma$ allowed uncertainties (see Table~\ref{tab:fitCOS}). The inferred $b$-parameter for \SiII\ is somewhat higher than those for \OI\ and \NI. This is possibly due to the presence of multiple components in the strong \SiII\ transitions. The $b$-parameter for \SII\ is not well constrained since all three transitions are on the linear part of the COG. It is interesting to note that at least one line in each case falls on the linear part of the COG, ensuring that the column densities we obtain using COG are robust.   

\begin{table} 
\begin{center}  
\caption{Component-by-component Voigt Profile Fit Parameters for the Metal Lines Detected 
in the High-resolution Keck Spectrum.}       
\begin{tabular}{ccccc}    
\hline \hline  
Ion    &           \zabs\          &   $b$ (\kms)  &  $\log N/$\sqcm &  $T$(K)$^{1}$  \\            
\hline 
\MgI   &   0.429616$\pm$0.0000013  &   5.9$\pm$0.4 & 10.82$\pm$0.16  & $10^{4}$  \\ 
\MgII  &          ...              &   5.9$\pm$0.0 & 12.28$\pm$0.02  & ...       \\ 
\FeII  &          ...              &   5.6$\pm$0.0 & 12.34$\pm$0.03  & ...       \\ 
\\                                                                       
\MgI   &   0.429685$\pm$0.0000015  &   5.0$\pm$0.5 & 10.76$\pm$0.17  & $10^{4}$  \\ 
\MgII  &          ...              &   5.0$\pm$0.0 & 12.10$\pm$0.03  & ...       \\ 
\FeII  &          ...              &   4.6$\pm$0.0 & 12.31$\pm$0.03  & ...       \\ 
\\                                                                      
\MgI   &   0.429805$\pm$0.0000009  &   6.3$\pm$0.2 & 12.08$\pm$0.02  & 156$^{a}$       \\ 
\MgII  &          ...              &   6.3$\pm$0.0 & 13.91$\pm$0.04  & ...       \\ 
\FeII  &          ...              &   6.3$\pm$0.0 & 13.68$\pm$0.02  & ...       \\ 
\CaII  &          ...              &   6.3$\pm$0.0 & 11.90$\pm$0.02  & ...       \\ 
\MnII  &          ...              &   6.3$\pm$0.0 & 11.72$\pm$0.09  & ...       \\ 
\\                                                                      
\MgI   &   0.429862$\pm$0.0000011  &   4.4$\pm$0.2 & 11.40$\pm$0.06  & $10^{4}$  \\ 
\MgII  &          ...              &   4.4$\pm$0.0 & 13.73$\pm$0.05  & ...       \\ 
\FeII  &          ...              &   4.0$\pm$0.0 & 13.49$\pm$0.02  & ...       \\ 
\CaII  &          ...              &   4.1$\pm$0.0 & 11.22$\pm$0.07  & ...       \\ 
\MnII  &          ...              &   4.0$\pm$0.0 & 11.20$\pm$0.26  & ...       \\ 
\\                                                                        
\MgII  &   0.430245$\pm$0.0000014  &   1.9$\pm$0.8 & 11.75$\pm$0.04  & ...$^{b}$  \\ 
\FeII  &          ...              &       ...     &  $<$11.00       & \\     
\MgI   &          ...              &       ...     &  $<$10.60       & \\    
\hline   
\label{tab:fitHIRES}      
\end{tabular} 
\end{center} 
\vskip-0.3cm  
Notes-- $^{1}$Assumed gas temperature in the Voigt profile model.   
$^{a}$Constrained from the $T_{01}$ measured from the column densities of $J=$0 and $J=$1 levels of \HH.  
$^{b}$Not assumed since no ions other than \MgII\ are detected. 
\end{table}   

\begin{table*} 
\begin{center} 
\caption{Voigt profile fit parameters of \HH.}        
\begin{tabular}{ccccccccccc}    
\hline \hline  
\zabs &   $b(\HH)$ &  \multicolumn{5}{c}{$\log N(\HH)$} & $\log N(\HH)_{\rm tot}$ & $\log f_{\hh}$  &  $T_{01}$  &   $\chi^2_{\rm red}$ \\ \cline{3-7} 
      &   (\kms)   &    $J= 0$  &  $J= 1$   &   $J=2$    &  $J=3$   &  $J=4^{}$ &  (cm$^{-2}$)      &                 &   (K)   & \\ \cline{3-7}  
  (1)  &  (2)      &            &           &  (3)       &          &     & (4)    &  (5)   &  (6)   &  (7)   \\  
\hline  
0.429807 & 7.1$\pm$0.3  &  15.74$\pm$0.08   &  16.22$\pm$0.09   &  15.23$\pm$0.03  & 14.83$\pm$0.03 & $<$13.8   & 16.36$\pm$0.08 & $-$2.84$\pm$0.17 & 156$\pm$47  & 2.1 \\  
0.429807 & 2.8$\pm$0.3  &  17.65$\pm$0.04   &  18.09$\pm$0.02   &  17.23$\pm$0.07  & 15.93$\pm$0.15 & $<$13.5   & 18.27$\pm$0.03 & $-$0.98$\pm$0.14 & 143$\pm$17  & 2.0 \\  
\hline 
\label{tab:H2fit} 
\end{tabular} 
\end{center} 
\vskip-0.3cm  
Notes-- (1) Median absorption redshift. All \HH\ absorption lines from different $J$ levels are consistent within $\pm$5~\kms\ of the median redshift. (2) Doppler parameters obtained from Voigt profile fitting. Note that {\sc vpfit} converges at two different solutions depending on the initial $b(\HH)$ values (see text). (3) \HH\ column densities for different $J$ levels. Standard 3$\sigma$ upper limits, obtained from the nondetection of the $\rm L10R4$ transition ($f\lambda=$30.76), are provided for the $J=$~4 level. Limits are calculated assuming that the $b$-parameter of the undetected $\rm L10R4$ line is the same as the corresponding $b(\HH)$ listed in column~2. (4) Total \HH\ column density, i.e., the sum of column densities obtained from different $J$ levels. (5) Logarithmic molecular fraction. (6) Rotational excitation temperature. (7) Reduced $\chi^2$ value for the fit as returned by {\sc vpfit}.           
\end{table*}   

The AOD technique \citep[]{Savage91} is known to provide accurate column density measurements for unsaturated absorption lines. For saturated absorption the AOD method yields a lower limit on the column density. The column densities from AOD measurements are also listed in Table~\ref{tab:summaryN}. Note that AOD column densities are consistent, within 1$\sigma$ measurement uncertainties, with both {\sc vpfit} and COG column densities, whenever available. For our ionization modeling in the next section we adopt the column densities that are listed in the last column of Table~\ref{tab:summaryN}. For a given ion, the adopted column density is simply the average of total column densities we obtained using the different methods (i.e. {\sc vpfit}, COG, and AOD).

\begin{table} 
\begin{center}  
\caption{Single-component Voigt Profile Fit Parameters for the 
Metal Lines Detected in the Medium-resolution COS Spectra.}       
\begin{tabular}{cccc}    
\hline \hline  
Ion    &           \zabs\          &   $b$ (\kms)  &  $\log N/$\sqcm   \\             
\hline 
\NI    &  0.429815$\pm$0.000003  &    6.8$\pm$0.6  &  14.78$\pm$0.06   \\ 
\OI    &  0.429811$\pm$0.000003  &    7.7$\pm$0.3  &  15.87$\pm$0.06   \\ 	 
\ArI   &  0.429828$\pm$0.000010  &    6.3$\pm$0.0  &  13.14$\pm$0.08   \\ 
\CI    &  0.429805$\pm$0.000000  &    6.3$\pm$0.0  &  13.88$\pm$0.13$^{a}$   \\ 
\SII   &  0.429818$\pm$0.000012  &   16.8$\pm$4.0  &  14.33$\pm$0.06   \\   	
\hline   
\label{tab:fitCOS}      
\end{tabular} 
\end{center}  
\vskip-0.3cm  
Notes-- Zero error indicates that the corresponding parameter was kept  
tied$/$fixed while fitting.   
$^{a}$This value should be treated as an upper limit. 
\end{table}   

\subsection{Ionization Models}      
\label{sec:mod}   

In this section we explore the chemical$/$physical conditions of the cool gas-phase, giving rise to a plethora of \HH\ and neutral$/$singly ionized metal absorption lines, using the photoionization simulation code {\sc cloudy} \citep[v13.03, last described by][]{Ferland13}. We use the observed total column densities of different species to constrain the model parameters and assume that all the $N(\HI)$ is associated with the \HH\ component \citep[but see][]{Srianand12}. In order to obtain an accurate \HH\ equilibrium abundance, we use full $J$ resolved calculations, as described in \cite{Shaw05}, using the {\sc `atom \HH'} command. The ionizing radiation used in our model is the extragalactic UV background radiation at $z=$~0.42 contributed by both QSOs and galaxies \citep[]{Haardt12}. In addition, cosmic rays are added with an ionization rate of $\log \Gamma_{\rm CR}=-$17.3 \citep[]{Williams98}. We do not consider the effect of a galaxy$/$stellar radiation field since we do not know the SFR of the host galaxy. However, we note that the effect of a galaxy radiation field will be negligible at this redshift for an impact parameter of $\sim$50~kpc from a sub-$L_{\ast}$ ($\sim$0.5$L_{\ast}$) host galaxy \citep[see, e.g.][]{Narayanan10,Werk14}. The lack of higher-$J$ (i.e. $J>3$) excitations also suggests that the prevailing radiation field is not strong \citep[]{Shull82,Tumlinson02}. The absorbing gas is assumed to be a plane-parallel slab irradiated by the ionizing radiation from one side.      

\begin{table} 
\begin{center} 
\caption{Summary of total column density measurements.}          
\begin{tabular}{ccccc}    
\hline \hline  
Ion       &    {\sc vpfit}      &      AOD$^{a}$    &     COG           &   Adopted Value$^{b}$  \\  
\hline  
\OI\      &   15.87$\pm$0.06    &  15.90$\pm$0.19   &  16.02$\pm$0.10   &   15.93$\pm$0.12    \\  
\SII\     &   14.33$\pm$0.06    &  14.21$\pm$0.20   &  14.25$\pm$0.25   &   14.26$\pm$0.16    \\  
\SiII\    &   ...               &  14.61$\pm$0.30   &  14.70$\pm$0.09   &   14.66$\pm$0.18    \\  
\CI\      &   $<$13.9           &  ...              &  ...              &   $<$13.9           \\       	     
\NI\      &   14.78$\pm$0.06	&  14.55$\pm$0.19   &  14.58$\pm$0.06   &   14.65$\pm$0.09    \\   
\ArI\     &   13.14$\pm$0.08    &  12.99$\pm$0.28   &  ...              &   13.10$\pm$0.16    \\    
\MgI\     &    12.20$\pm$0.04   &  12.20$\pm$0.14   &  ...              &   12.20$\pm$0.09    \\  
\MgII\    &    14.14$\pm$0.04   &  $>$13.7          &  ...              &   14.14$\pm$0.04    \\  
\FeII\    &    13.92$\pm$0.02   &  13.90$\pm$0.09   &  ...              &   13.91$\pm$0.05    \\  
\MnII\    &    11.83$\pm$0.13   &  11.93$\pm$0.64   &  ...              &   11.88$\pm$0.41    \\   
\CaII\    &    11.98$\pm$0.03   &  11.97$\pm$0.17   &  ...              &   11.98$\pm$0.10    \\  
\hline 
\label{tab:summaryN}  
\end{tabular} 
\end{center} 
\vskip-0.3cm  
Notes-- Measurement is not taken using the particular method when empty.     
$^{a}$Except for the \SII\ and \FeII\, the weakest available transitions are 
used for estimating AOD column densities. The weakest transitions of \SII\ 
($\lambda$1250 and $\lambda$1253) and \FeII\ ($\lambda$2374) are noisy and hence are not used.  
$^{b}$Adopted column densities for our photoionization model.      
\end{table}   

\begin{figure} 
\centerline{
\vbox{
\centerline{\hbox{ 
\includegraphics[width=0.50\textwidth]{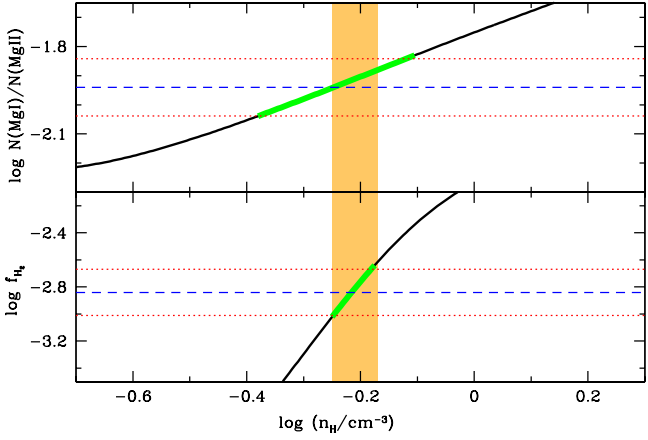} 
}} 
}}  
\caption{Photoionization-model-predicted molecular fraction (bottom) and  \MgI-to-\MgII\ column density ratio (top) as a function of gas density. In each panel the horizontal dashed and dotted lines represent the corresponding observed value and its 1$\sigma$ uncertainty, respectively. The vertical band indicates the allowed density range within which the model-predicted values of both $\log f_{\hh}$ and $\log N(\MgI)/N(\MgII)$  match with the observations. We adopt $\log n_{\rm H} = -$0.2 as our density solution.     
}  
\label{fig:model1}   
\end{figure} 

The total \HI\ column density, i.e., $\log N(\HI) = 19.50\pm0.15$, measured in this sub-DLA indicates that the absorbing gas is optically thick. Therefore, unlike the optically thin medium, the ionization corrections for different species are no longer independent of the assumed metallicity and$/$or $N(\HI)$. Moreover, the dust-to-gas ratio is another important input model parameter since \HH\ is present. The observed total $N(\OI)$ corresponds to an uncorrected gas-phase metallicity of $\rm [O/H]=-$0.26$\pm$0.19. Thus, we use an input metallicity of $\rm [X/H] = -0.3$ dex for our model. Assuming the intrinsic $\rm [Fe/X]$ to be solar, the dust-to-gas ratio, relative to the solar neighborhood, is defined as   
\begin{equation} 
\kappa = 10^{[\rm {X/H}]}(1-10^{[\rm {Fe/X}]})~,  
\end{equation}  
where $\rm X$ is an element that does not deplete onto dust significantly \citep[see][]{Wolfe03}. Assuming $\rm X\equiv S$, we obtain $\log \kappa = -0.45$. Hence, we input $\log \kappa =-$0.5, with dust composition similar to Milky Way, in our model. The simulation is run on a density grid. For each density, the calculation is stopped when $N(\HI)$ in the simulation matches the observed value.  

The variations of $\log f_{\hh}$ and the $N(\MgI)/N(\MgII)$ ratio with density in our model are shown in the bottom and top panels of Figure~\ref{fig:model1}, respectively. Parameter $f_{\hh}$ shows a sharp increase with density. The observed $f_{\hh}$ value and its 1$\sigma$ uncertainty correspond to a density in the range $\log n_{\rm H}\simeq-$0.15 to $-$0.25. Interestingly, the model-predicted $N(\MgI)/N(\MgII)$ ratios also match the observed value within 1$\sigma$ allowed uncertainty in this density range. We	therefore adopt $\log n_{\rm H}=-$0.2, corresponding to an ionization parameter of $\log U=-$5.6, as the density solution. The predicted gas temperature at this density (i.e. 135~K) is also consistent with the observed $T_{01}$.     

\begin{figure} 
\centerline{
\vbox{
\centerline{\hbox{ 
\includegraphics[width=0.50\textwidth]{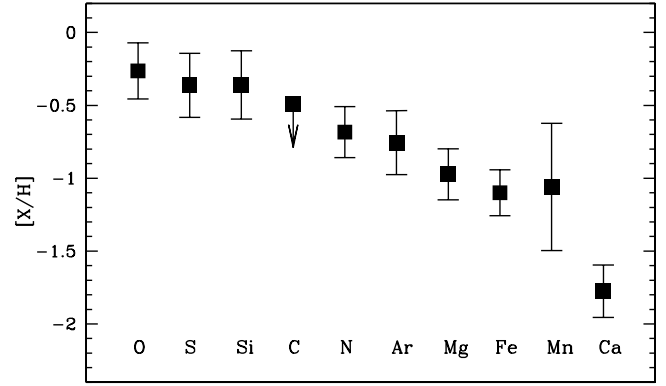}  
}} 
}}  
\caption{Ionization-corrected abundances of different elements for our adopted photoionization model. Note that the $x$-axis here does not represent anything. We find that Mg, Fe, Mn, and Ca are significantly depleted.}         
\label{fig:abundance}    
\end{figure}

We use {\sc cloudy}-computed ionization fractions of \HI\ and different metal ions, at $\log n_{\rm H}=-$0.2, for estimating ionization-corrected abundances. Abundances of different metals, after ionization corrections, are shown in Figure~\ref{fig:abundance} and summarized in Table~\ref{tab:abundances}. Abundances are calculated using the standard convention: $\rm [X/H]= \log (X/H)_{gas} - \log (X/H)_{\odot}$. Ionization correction, in this context, is defined as $\epsilon_{\rm X}= \log (f_{\hi}/f_{{\rm X}i})$, where, $f_{{\rm X}i}$ and $f_{\hi}$ are the ionization fractions of the ($i-1$)th ionization state of element $\rm X$ and \HI, respectively. It is apparent from Table~\ref{tab:abundances}, that with the exception of \OI\ and \NI, the ionization correction is important (i.e., $|\epsilon_{\rm X}|>$0.1 dex) for all neutral species. On the other hand, except for \CaII, ionization corrections are negligibly small for all singly ionized species. The ionization-corrected abundance of oxygen $\rm [O/H]=-0.26\pm0.19$ is the same as the uncorrected value. Sulfur and silicon have 0.1 dex lower abundances as compared to oxygen. Nevertheless, the abundances of O, S, and Si are consistent within 1$\sigma$. Estimated $\rm [C/O]<-0.23$ and $\rm [N/O]=-0.42\pm0.25$ suggest that both C and N are underabundant with respect to O. Additionally, we measure $\rm [Ar/H]=-0.76\pm0.22$ dex. Similar abundance for the ``inert gas" Ar is seen in the Milky Way ``cool clouds" \citep[]{Savage96a}. A detailed discussion on elemental abundances and depletions is presented in Section~\ref{sec:abundances}.                

\begin{table} 
\begin{center}  
\caption{Abundances of different elements.}         
\begin{tabular}{ccccc}    
\hline \hline  
Element & Ion &   $\rm [X/H]_{\rm uncorr}^{a}$  &  $\epsilon_{\rm X}^{b}$  & $\rm[X/H]_{\rm corr}^{c}$  \\             
\hline 
O       &  \OI    &   $-$0.26$\pm$0.19  &   $-$0.003     &   $-$0.26$\pm$0.19  \\ 
S       &  \SII   &   $-$0.36$\pm$0.22  &   $-$0.002     &   $-$0.36$\pm$0.22  \\ 
Si      &  \SiII  &   $-$0.35$\pm$0.23  &   $-$0.009     &   $-$0.36$\pm$0.23  \\  
C       &  \CI    &   $-$2.03           &   $+$1.542     &   $<-$0.49          \\ 
N       &  \NI    &   $-$0.68$\pm$0.17  &   $-$0.004     &   $-$0.68$\pm$0.17  \\ 
Ar      &  \ArI   &   $+$0.04$\pm$0.22  &   $-$0.800     &   $-$0.76$\pm$0.22  \\  
Mg      &  \MgI   &   $-$2.90$\pm$0.17  &   $+$1.927     &   $-$0.97$\pm$0.17  \\ 
Fe      &  \FeII  &   $-$1.09$\pm$0.16  &   $-$0.010     &   $-$1.10$\pm$0.16  \\ 
Mn      &  \MnII  &   $-$1.05$\pm$0.44  &   $-$0.009     &   $-$1.06$\pm$0.44  \\ 
Ca      &  \CaII  &   $-$1.86$\pm$0.18  &   $+$0.085     &   $-$1.78$\pm$0.18  \\ 
\hline   
\label{tab:abundances}        
\end{tabular} 
\end{center}  
\vskip-0.3cm  
Notes-- $^{a}$Abundance without ionization correction.  
$^{b}$Ionization correction as defined in the text.      
$^{c}$Abundance after ionization correction.  
\end{table}   

\begin{figure*}  
\centerline{
\vbox{
\centerline{\hbox{ 
\includegraphics[width=0.95\textwidth]{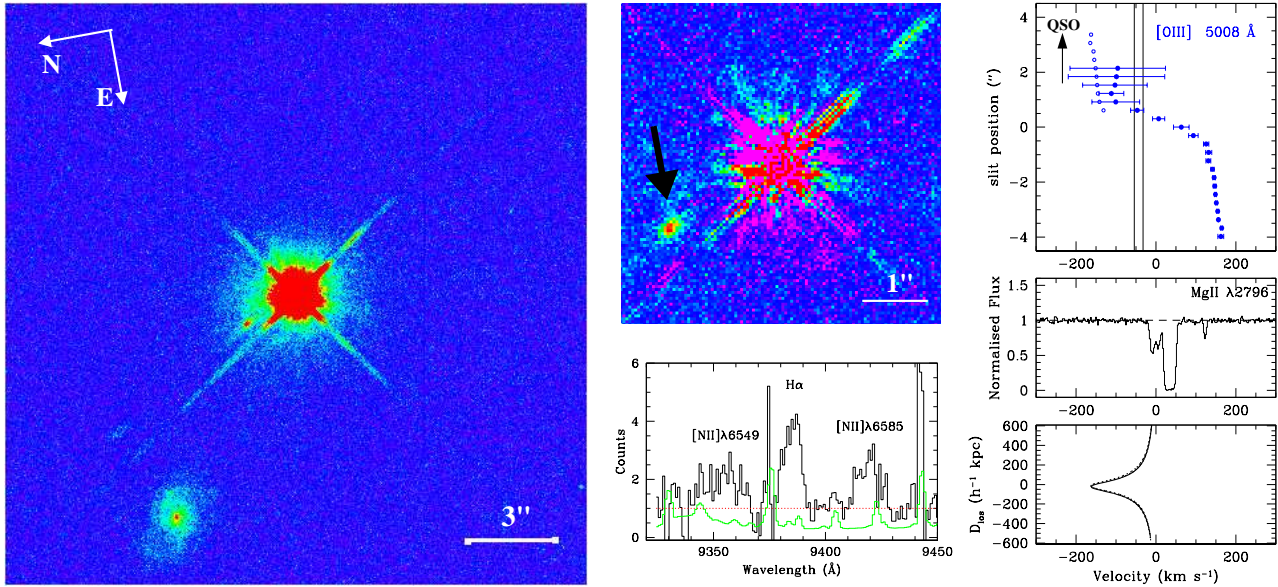} 
}} 
}}  
\caption{{\it Left:} $HST/$WFPC2 image of the QSO field. The host galaxy is located toward the northeast with respect to the QSO at the center. {\it Middle:} H$\alpha$ and \NII\ emission lines from the host galaxy are shown in the bottom panel; the presence of another galaxy, possibly a dwarf-satellite of the host galaxy, is clearly visible in the QSO PSF-subtracted image of the field in the top panel. {\it Right:} The rotation curve, \MgII\ absorption kinematics, and the halo model of \cite{Steidel02}. The halo model is inconsistent with the \MgII\ kinematics for the most part (see text). The velocities below $-50$~\kms\ in the rotation curve are affected by a skyline contamination. The open circles in the rotation curve are the mirror of the corresponding measurements at $+ve$ velocities.}            
\label{fig:HSTimage}    
\end{figure*} 

Although the ionization model presented above is based on the observed total column densities, the dominating component is the one at $\sim$0~\kms\ in which \HH\ is present. This component comprises about $\sim$60\% of the total \MgII\ and \FeII, and $\sim$75\% of the total \MgI\ column densities. Thus, the model parameters essentially correspond to this component. Due to the lack of $N(\HI)$ information in individual components, we could not build component-by-component models. This is because all the higher-order Lyman series lines are heavily saturated. Nonetheless, from the measured $N(\MgI)/N(\MgII)$ and $N(\MgII)/N(\FeII)$ ratios we infer that the density in the two weak components at $v\sim-$30~\kms\ (see Figure~\ref{fig:fitHIRES}) could be as high as seen in the \HH\ component, provided that the metallicities are not very low. The weakest \MgII\ component at $v\sim$90~\kms\ is more likely to be similar to the ``weak systems" studied by \cite{Rigby02}.  

If we use the higher $N(\HH)$ solution corresponding to $\log f_{\hh}=-$0.98 then the required density is $\log n_{\rm H} =$~0.8, i.e., an order of magnitude higher than our adopted density. At such a high density the $N(\MgI)/N(\MgII)$ ratio is a factor of $\sim$5 higher than the observed value. Additionally, the predicted photoionization equilibrium temperature at this density (i.e. 40~K) is significantly lower than the corresponding $T_{01}$ value (143$\pm$17~K; see Table~\ref{tab:H2fit}). Next, \cite{Srianand12}  have found that \HH\ components in high-$z$ DLAs are compact (with sizes of $\lesssim$15~pc) and contain only a small fraction ($\lesssim$10\%) of the total $N(\HI)$. If this is also true for the present system, then the \HH-bearing gas would require a much higher density ($>10$~cm$^{-3}$), which is inconsistent with the density range allowed by the $N(\MgI)/N(\MgII)$ ratio (see Figure~\ref{fig:model1}). Finally, we note that an appreciable stellar contribution to the ionizing background on top of the extragalactic UV background would require a higher gas density in order to reproduce the same amount of $N(\HH)$ for a given $N(\HI)$ \citep[see, e.g.,][]{Dutta15,Muzahid15a}.

\section{Galaxy Analysis}  
\label{sec:galana}  

The $HST/$WFPC2 F702W image of the field is shown in the left column of Figure~\ref{fig:HSTimage}. The host galaxy is located at an impact parameter of $\rho=$~48.4~kpc toward the northeast with respect to the QSO. The host galaxy has an inclination angle of $i=48.3^{+3.5}_{-3.7}$~deg. The angle between the projected major axis of the host galaxy and the QSO sightline (i.e. the azimuthal angle, $\Phi$) is $14.9^{+0.6}_{-4.9}$~deg. The azimuthal angle indicates that the absorbing gas is located near the projected major axis. The $B-K$ color of 2.06 \cite[]{Nielsen13a}, suggests a moderate star formation rate (SFR) in the host galaxy. Due to the unavailability of flux-calibrated spectra (see Section~\ref{sec:galdata}), we could not constrain the SFR of the host galaxy. Nonetheless, the presence of H$\alpha$ and [\NII] emission lines (see Figure~\ref{fig:HSTimage}), allowed us to measure a redshift of \zgal~$=$~0.42966$\pm$0.00016 and a metallicity of $\rm 12+\log (O/H)=8.68\pm0.09$ for the host galaxy. The metallicity is estimated using the $N2$ relation of \citet{Pettini04}, i.e.,  $\rm 12+\log (O/H)= 8.90+ 0.57\times N2$, where $N2\equiv$ \NII$/\rm H\alpha$. We note that (i) the \zgal\ is consistent with the \zabs\ within 1$\sigma$ uncertainty and (ii) the host galaxy has a metallicity consistent with the solar value \citep[i.e., $\rm 12+\log (O/H)=8.69$,][]{Asplund09}. 

The point spread function (PSF) subtracted and magnified image of the QSO field is shown in the middle column of Figure~\ref{fig:HSTimage}. Interestingly, we found another galaxy at a much lower impact parameter ($\rho=2.06''$) in this image. In fact, the presence of this dwarf galaxy is noticeable even in the left column of Figure~\ref{fig:HSTimage}. This galaxy could be a satellite of the bigger galaxy at $\rho=$~48.4~kpc. However, we do not have a spectrum of this galaxy to constrain its redshift. Assuming that the dwarf galaxy is at \zabs, we obtain an impact parameter of $\rho =11.6$~kpc and an apparent magnitude of $m_{\rm F702W}=23.20\pm0.06$ (Vega). The dwarf-satellite galaxy is highly inclined with $i=70.0^{+15.0}_{-21.1}$~deg. Similar to the bigger galaxy, the projected major axis of the satellite galaxy also has a small azimuthal angle, i.e., $\Phi=10.6^{+19.4}_{-10.6}$~deg, with respect to the QSO sightline.      

The rotation curve of the spectroscopically confirmed host galaxy is constructed using the [\OIII]~$\lambda$5008 emission line and is shown in the right column of Figure~\ref{fig:HSTimage}. The zero velocity here corresponds to the \zgal\ and not to \zabs. The galaxy has a maximum projected rotation velocity of $|v_{\rm max}|\sim$170~\kms. Note that the rotation curve at the $-ve$ velocities is affected by a skyline contamination to the [\OIII]~$\lambda$5008 line. 

We now construct the halo model of \cite{Steidel02} using $i$, $\Phi$, $\rho$, and $v_{\rm max}$ of the host galaxy. Following \cite{Kacprzak10a} we use a gas scale height, $h_{v}$, a free parameter of the model, of 1000~kpc (for a non-lagging halo above the galaxy plane). Such a model examines the range of allowed velocities along the line of sight that are consistent with extended galaxy rotation. The model is shown in the bottom panel of the right column of Figure~\ref{fig:HSTimage}. The $D_{\rm los}$ is the distance along the line of sight relative to the point where the sightline intercepts the projected mid-plane of the galaxy. A halo model is considered successful when the $D_{\rm los}$ curve spans the same velocity range as that of the observed \MgII\ absorption. It is important to notice that the observed \MgII\ kinematics is {\it not} consistent with such a model of a co-rotating disk but is consistent with a counter-rotating lagging halo.          

\begin{figure} 
\centerline{
\vbox{
\centerline{\hbox{ 
\includegraphics[width=0.50\textwidth]{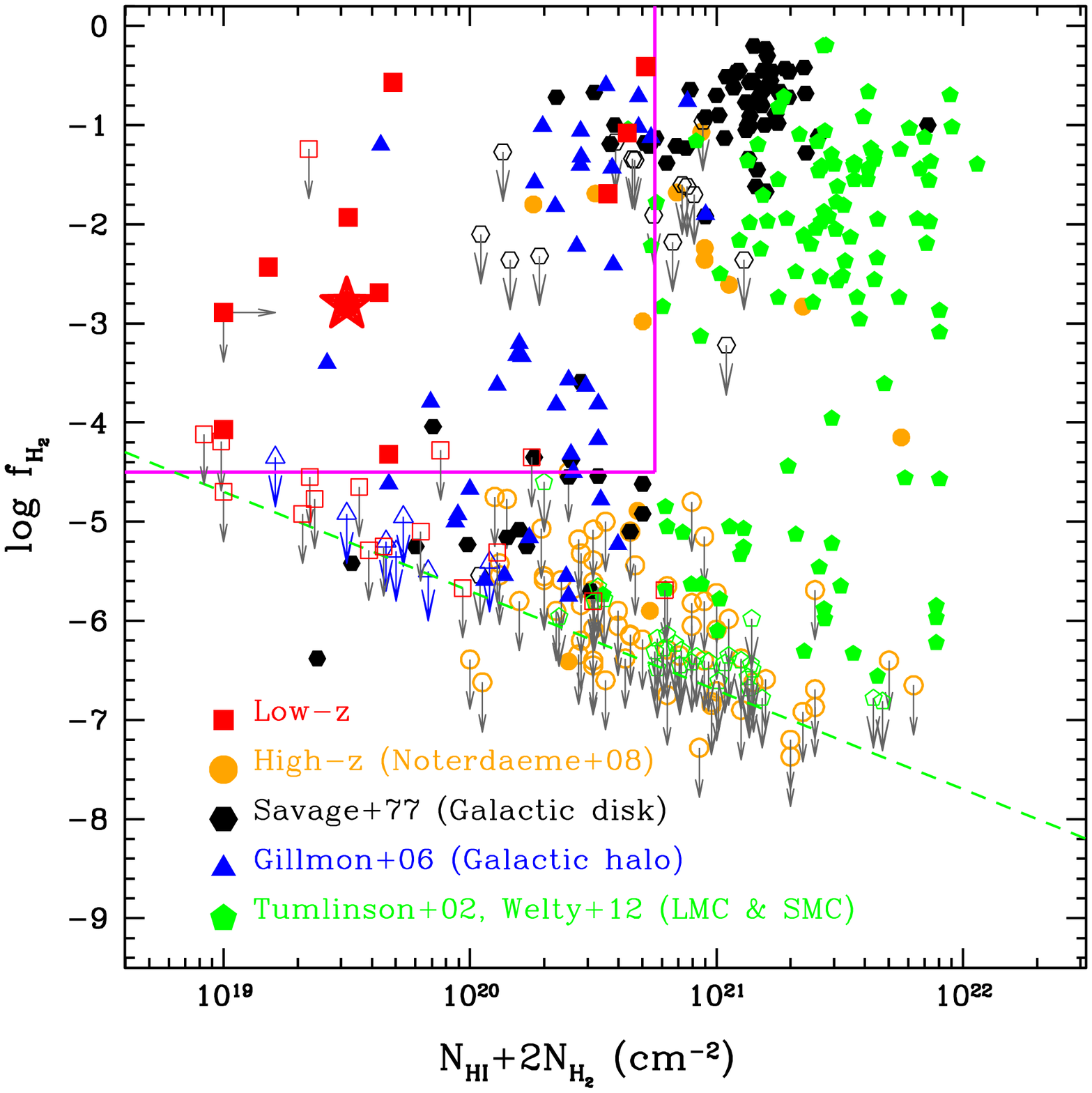} 
}} 
}}  
\caption{Molecular fraction versus the total hydrogen column density, $N(\HI+\HH)$, for different astrophysical environments: low-$z$ \citep[]{Muzahid15a} and high-$z$ \citep[]{Noterdaeme08} DLAs$/$sub-DLAs, the Galactic disk \citep[]{Savage77} and halo \citep[]{Gillmon06}, and the Magellanic Clouds \citep[]{Tumlinson02,Welty12}. For each sample, systems with detected \HH\ are shown by filled symbols and the limits are shown by open symbols. The system studied here is shown by the red star. The magenta lines delineate the region ($\log f_{\hh}>-$4.5 and $\log N(\HI+\HH)<$~20.7) within which all the low-$z$ \HH\ systems and a majority (i.e., $\sim$65\%) of the Galactic halo systems reside. The dashed line corresponds to $\log N(\HH)=$~14.0, which can be treated as a typical detection threshold.          
}  
\label{fig:fH2-NH}   
\end{figure} 

\section{Discussion} 
\label{sec:diss}

\subsection{Physical conditions}

As mentioned in Section~\ref{sec:intro}, the Lyman- and Werner-band absorption of \HH\ is commonly observed in a wide variety of astrophysical environments. In Figure~\ref{fig:fH2-NH} we have summarized the \HH\ observations from the literature. It is evident from the figure that the system studied here, consistent with other low-$z$ \HH\ systems, shows an unusually large $f_{\hh}$ for its total (atomic+molecular) hydrogen column density, $N_{\rm H}$. As noted by \cite{Muzahid15a}, the low-$z$ \HH\ systems populate the upper left corner of the $f_{\hh}$--$N_{\rm H}$ plane, as marked by the magenta lines in the plot. \HH\ systems in the Galactic disk \citep[]{Savage77}, Magellanic Clouds \citep[]{Tumlinson02,Welty12}, and high-$z$ samples \citep[]{Noterdaeme08} are mostly DLAs with $\log N_{\rm H}>$~20.7. Additionally, the systems with $\log N_{\rm H}<$~20.7 either do not show \HH\ absorption or have $\log f_{\hh}<-4.5$. In contrast, a significant fraction ($\sim$65\%) of the Galactic halo systems \citep[]{Gillmon06}, with detected \HH, have $f_{\hh}$ and $N(\HH)$ values in the ranges similar to those seen for the low-$z$ systems. The physical conditions in the \HH-absorbing gas, therefore, are more similar to those found for diffuse molecular clouds in the Galactic halo than to those for molecular clouds in the Galactic disk.   

\begin{figure} 
\centerline{
\vbox{
\centerline{\hbox{ 
\includegraphics[width=0.50\textwidth]{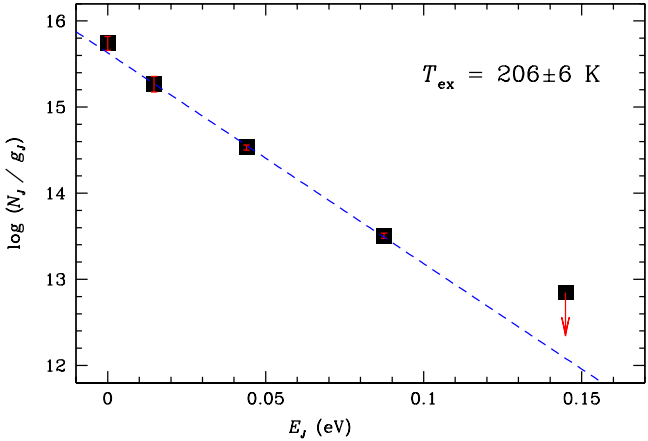} 
}} 
}} 
\vskip-0.2cm  
\caption{Excitation diagram for the rotational level populations of \HH. The population ratios of different rotational levels are expressed as a Boltzmann distribution [$N(J)\propto\exp(-E_{J}/kT)$]. Interestingly, a single excitation temperature can explain all the level populations.     	  
}  
\label{fig:Tex}     
\end{figure}

The relative level populations of \HH\ provide a sensitive diagnostic of gas temperature \citep[]{Tumlinson02,Srianand05}. The excitation diagram for this system, shown in Figure~\ref{fig:Tex}, indicates that all the $J$-levels are consistent with a single excitation temperature of $T_{\rm ex}=206\pm6$~K. Note that the $T_{\rm ex}$ is consistent with the $T_{01}$ within the 1$\sigma$ allowed range. Such gas temperatures, along with the observed $N(\HI)$ and $f_{\hh}$ values, are consistent with the diffuse atomic gas-phase of the Galactic ISM as characterized by \cite{Snow06}.    

When \HH\ is sufficiently shelf-shielded and collisional processes dominate the $J=0$ and 1 level populations, $T_{01}$ represents the kinetic temperature of the gas \citep[]{Roy06,Snow06}. Using \HI\ 21 cm observations, \cite{Roy06} have shown that the \HI\ spin temperatures, $T_s$, are correlated with the $T_{01}$ for systems with $\log N(\HH)\gtrsim$16.0. Note that \cite{Kanekar03} observed but did not detect \HI\ 21 cm absorption from this sub-DLA. From the nondetection, the authors placed a 3$\sigma$ lower limit on the spin temperature of $T_s>$~980~K, assuming a covering factor of $f_c\sim1$. Clearly, such a high temperature is not consistent with our $T_{\rm ex}$ or $T_{01}$ measurements. The apparent mismatch between $T_s$ and $T_{\rm ex}$ (or $T_{01}$) could be due to a significantly lower covering factor of the \HH-bearing gas-phase. Dense and compact structures of \HH-absorbing gas with small covering fraction are commonly observed \citep[e.g.,][]{Srianand12,Dutta15} and are predicted in numerical simulations as well \citep[e.g.,][]{Hirashita03}.    

The populations of higher rotational levels (i.e., $J>3$) are determined by UV pumping and formation pumping \citep[]{Shull82,Tumlinson02}. This system, consistent with other low-$z$ \HH\ absorbers, do not show absorption from $J>3$ levels. The lack of high-$J$ excitations clearly indicates an absence of a local UV radiation field. The large impact parameter ($\rho\sim50$~kpc) of the identified host galaxy is consistent with such an observation. 

\subsection{Elemental abundances and dust}  
\label{sec:abundances}      

Measurements of elemental abundances in DLAs and sub-DLAs provide essential clues on cosmic history of gas and dust in host galaxies. The main challenge in deriving abundance, particularly in metal-rich systems, is to disentangle the nucleosynthetic contributions from dust depletion effects. Assessment of intrinsic abundance, from gas-phase abundance, of an element becomes complicated when a fraction of that element depletes onto dust. A non-negligible amount of dust with a dust-to-gas ratio of $\log \kappa=-0.45$ is present in the sub-DLA studied here. Nevertheless, we could estimate gas-phase abundances of several $\alpha$-elements (i.e., O, S, Si, Ar, Mg, and Ca) and Fe-peak elements (i.e., Fe and Mn), as shown  \linebreak in Figure~\ref{fig:abundance}.             

In general, metallicity is derived from $\rm [O/H]$ due to the charge transfer reaction between \HI\ and \OI. The inferred metallicity from our photoionization model is $\rm [O/H]=-0.26\pm0.19$. \cite{Petitjean06} noted a correlation between gas-phase metallicity and \HH\ detection for high-$z$ DLAs in which higher-metallicity DLAs tend to show \HH\ more often than lower-metallicity systems \citep[see also][]{Noterdaeme08}. Furthermore, \cite{Ledoux03} and \cite{Noterdaeme08} found that the \HH\ detection is more frequent for systems with higher $\log \kappa$ and $N_{\rm Fe}^{\rm dust}$ values, where the latter is the column density of dust in Fe. Note that all \HH\ absorbers at high-$z$ have $\log N_{\rm Fe}^{\rm dust}>14.7$ and $\log \kappa>-1.5$. These observations are understood in view of efficient formation of \HH\ on the surface of dust grains. Assuming the intrinsic $\rm [Fe/S]$ to be solar, a $\log \kappa$ of $-0.45$ corresponds to $N_{\rm Fe}^{\rm dust} \sim 10^{14.6}$~\sqcm\ for the present system. Clearly, the detection of \HH\ is consistent with the overall trends seen in high-$z$ DLAs. Most importantly, the inferred high metallicity is inconsistent with a scenario of gas accretion since the accreting gas is typically metal poor with $\rm [X/H]<-1.0$ \citep[]{Ribaudo11,Kacprzak12b,Lehner13}.    

It is now well accepted that $\alpha$-elements are produced via Type II supernovae (SNe II) from core collapse of massive \linebreak ($>8M_{\odot}$) stars. Fe-peak elements, on the contrary, arise in SNe Ia with low-$/$intermediate-mass progenitors. Delayed evolution of low- and intermediate-mass stars leads to a characteristic time gap of $\sim$1 Gyr between formation of $\alpha$-elements and Fe-peak elements \cite[e.g.,][]{Hamann93}. Thus, the ratio of $\alpha$-elements to Fe-peak elements with similar condensation temperatures, i.e., $\rm [Mg/Fe]$, provides a sensitive diagnostic of star formation history. We found  $\rm [Mg/Fe] = 0.13\pm0.23$, i.e. consistent with the \linebreak solar value, for our system, indicating a lack of significant $\alpha$-enhancement. Moreover, $\rm [Mg/Mn] = 0.09\pm0.47$ is also consistent with the solar value. Therefore, it seems that Fe, Mg, and Mn all have similar gas-phase abundances of $\rm [X/H]\sim-1.0$ indicating a moderate depletion of $\gtrsim$0.7 dex. Similar depletion is seen in the Galactic ``warm diffuse clouds" \citep[]{Savage96a}. Additionally, the heavy depletion of Ca as we observed here (i.e. $>$1.5 dex) is also common in the Galactic diffuse ISM \citep[]{Welty99a}.  

Oxygen is almost entirely produced by SNe II, primarily during the central H-burning phase. Carbon can be produced in stars of all masses via He burning. The synthesis of N can arise from primary and$/$or secondary production via the CNO cycle. If N is synthesized from C produced in the core of the same star via He burning, it is called primary. Secondary N enrichment of the ISM occurs well after massive stars have undergone SNe II and seeded the ISM with oxygen. For ``primary N", the $\rm [N/O]$ ratio is $\sim-0.6$ dex for $\rm [O/H] < -0.3$. The $\rm [N/O]$ ratio increases with the $\rm [O/H]$ ratio for ``secondary N" \citep[]{Pettini08a}. The observed $\rm [N/O]= -0.42\pm0.25$ along with the near-solar metallicity suggest that the system lies near the ``knee" of the $\rm [N/\alpha]$ versus $\rm [\alpha/H]$ plot, and thus both primary and secondary channels could be responsible for the detected nitrogen \citep[see Figure~9 of][]{Pettini08a}. Finally, the underabundance of C in our system (i.e., $\rm [C/O] < -0.23$) is consistent with the measurements in Galactic halo stars \citep[see][]{Akerman04}.

\subsection{Comparison with local analogs}   

First, we discuss the remarkable similarity in physical conditions between the present system and the \HH\ absorber detected in HVC~$287.5+22.5+240$ probing the Leading Arm (LA) of the Magellanic Stream \cite[see][]{Sembach01}. The HVC is located at $\sim$50~kpc from the Sun, similar to the impact parameter of the host galaxy of the \zabs~$=$~0.4298 absorber. The HVC absorber shows $\log N(\HI)=19.90\pm0.05$, $\log N(\HH) = 16.80\pm0.10$, $\log f_{\hh} = -2.80\pm0.11$, and $T_{01}=133^{+37}_{-21}$~K. All these values are surprisingly similar to the system studied here. Moreover, as in the present case, \HH\ is detected only up to the $J=3$ level. In fact, none of the extragalactic \HH\ absorbers at low-$z$ show high-$J$ excitations \citep[see Table~2 of][]{Muzahid15a}. The abundance of sulfur ($\rm [S/H] = -0.60$) and dust-to-gas ratio ($\log \kappa = -0.66$), as measured in the HVC, are somewhat lower as compared to the present system. Based on the similarity in abundance pattern between the absorber and warm gas in the SMC, the authors concluded that \HH\ in the HVC is remnant material tidally stripped from the SMC. The large {\it in situ} \HH\ formation time ($\sim$1 Gyr) derived for the system further suggested that tidal stripping is the most relevant origin. 

Next, we discuss the \HH\ absorbers in the IVCs, with $|v_{\rm LSR}|=$~30--90~\kms, studied by \cite{Richter03}. The authors found a large sky coverage ($\sim$35\%) of IVCs with $N(\HI)>2\times10^{19}$~\sqcm\ and noted the ubiquity of a diffuse \HH\ component in the Galactic IVCs. For example, \HH\ is detected in 14 out of 29 (i.e., 48$\pm$13\%) IVC sightlines with spectra sensitive to detect \HH\ down to $\log f_{\hh}\sim-3$. Note that the \HH\ detection rate becomes 31$\pm$11\% for the low-$z$ DLA$/$sub-DLA sample of \cite{Muzahid15a} when we apply a similar $\log f_{\hh}$ cutoff. Thus, the \HH\ detection rate in low-$z$ DLAs$/$sub-DLAs is somewhat lower but consistent with the \HH\ detection rate in IVCs within the 1$\sigma$ Poisson error. Nevertheless, the molecular fractions of $\log f_{\hh}=$~$-$5.3 to $-$3.3 (with a median value of $-$4.3) obtained for the \HH-detected IVCs are significantly lower than those found for the extragalactic low-$z$ \HH\ absorbers with a median $\log f_{\hh}=-1.93\pm0.63$ \citep[see][]{Muzahid15a}. This is possibly due to the fact that the IVCs studied in the sample of \cite{Richter03} are located between 0.3--2.1~kpc from the plane of the Galactic disk. For the low-$z$ \HH\ absorbers, including the one studied here, the host galaxies are identified at $\rho>$10~kpc. Thus, albeit having near-solar to solar metallicities, the IVCs show lower molecular fractions, perhaps due to the higher radiation field in closer proximity to the Galaxy.

\subsection{Origin of sub-DLA and \HH}   

From the discussions above it is clear that the physical conditions and the chemical compositions of the \HH-bearing gas are comparable to the diffuse atomic gas-phase \citep[]{Snow06} which is consistent with the large impact parameter of the identified host galaxy. But how can molecules exist at $\sim$50~kpc from the host galaxy? In Section~\ref{sec:galana} we have shown that the observed \MgII\ kinematics is inconsistent with a co-rotating, extended disk. In addition, from the $B$-band absolute magnitude ($M_{B}= -20.35$) of the host galaxy we obtain the effective radius of the \HI\ disk of only $\sim$13~kpc using the scaling relation of \cite{Lah09}. Thus, the absorber is not stemming from an extended \HI\ disk.      

Molecular gas in a galaxy halo could arise either from (i) $in~situ$ formation \citep[]{Richter03} or from (ii) ejected disk material via a central starburst \citep[]{Geach14Nat} or through tidal$/$ram pressure stripping from a satellite galaxy \citep[]{Sembach01}. \HH\ formation takes place efficiently on the surface of dust grains if the gas is cool ($T\sim$100~K), dense, and mostly neutral \citep[]{Hollenbach71a,Shull82}. In such a case, the volume formation rate of \HH\ can be expressed as $Rn_{\rm H}$, where $R$ is the formation rate coefficient. $R\sim(1-3)\times10^{-17}$~cm$^{3}$~s$^{-1}$ in the Galactic ISM \citep[]{Jura74b}. For the range of physical conditions for the present system, the $in~situ$ \HH\ formation time is $t_f\sim(Rn_{\rm H})^{-1}\sim$5$\times10^{9}$~yr for $R=10^{-17}$~cm$^{3}$~s$^{-1}$ and $\log n_{\rm H}=-0.2$. The large $in~situ$ formation time suggests that the idea of ejected disk material is more suitable.  

Next, we discuss whether the absorbing gas is originating in an outflow from the host galaxy. Intervening absorption systems with high metallicities are generally thought to be tracing galactic-scale outflows \citep[]{Tripp11,Muzahid14,Muzahid15b}. Therefore, it is tempting to argue that the present system is probing a molecular outflow owing to its high gas-phase metallicity. Recent radio and millimeter observations have, indeed, detected large-scale molecular outflows \citep[e.g.,][]{Alatalo11,Geach14Nat}. The molecular outflow studied by \cite{Alatalo11} extends only up to $\sim$100 pc and is thought to be powered by an active galactic nucleus. However, the other molecular outflow, studied by \cite{Geach14Nat}, has a spatial extent up to $\sim$10~kpc and is driven by stellar feedback from a compact starburst galaxy with an SFR of 260~$M_{\odot}$~yr$^{-1}$. We note that the host galaxy of the present system is forming stars only at a moderate rate today. Besides, the azimuthal angle ($\sim$15\degree) suggests a large opening angle (2$\theta>150$\degree) for an outflow to intercept the QSO sightline. Note that an outflow from the dwarf-satellite candidate would require an opening angle of 2$\theta>160$\degree.   

Outflows from Seyfert galaxies typically have opening angles ranging between 60\degree--135\degree\ \citep[]{Hjelm96,Veilleux01,Mullar11}. In hydrodynamic simulations, the opening angles of outflows from star-forming galaxies are in the range 2$\theta\simeq$10\degree--45\degree\ near the base and 45\degree--100\degree\ above the disk \citep[see][and references therein]{Veilleux05}. Outflow opening angles above the disk of $\sim$100\degree\ are also confirmed via absorption-line studies \citep[]{Kacprzak12a,Kacprzak15,Bordoloi14c}. As noted by \cite{Veilleux05}, the induced Kelvin-Helmholtz instabilities, as the wind propagates through the ISM, lead to fragmentation and subsequent mixing of the wind material. Thus, at a large distance ($\sim$50~kpc) it is {\it not} surprising to detect wind material even at small azimuthal angles unless the flow is highly collimated.  

Finally, as mentioned in Section~\ref{sec:galana}, there is a dwarf galaxy at a distance of $\sim$12~kpc, provided that the galaxy has the same redshift as the absorber, from the QSO sightline. It is possible that the dwarf galaxy is interacting with the host galaxy gravitationally. In such a case, materials from the outer disk of the satellite galaxy can get stripped off due to tidal forces or due to ram pressure. As discussed in the previous section, the detection of \HH\ in the LA by \cite{Sembach01} is an excellent analogy. The Magellanic Clouds are located at $\sim$55~kpc from the MW, and the LA is thought to be stemming from tidal interaction between the Magellanic Clouds and the MW \citep[see][for a recent review]{DOnghia15}. Future spectroscopic observations of the dwarf galaxy are essential for further understanding of this intriguing \HH\ absorption system. \\      

\section{Summary}  
\label{sec:summ}  

We present a detailed analysis of an \HH-detected sub-DLA at \zabs~$=$~0.4298 in the spectra of QSO PKS~2128--123. Historically, the connection between a QSO absorber and an intervening galaxy was first demonstrated, observationally, by analyzing this absorber \cite[i.e.,][]{Bergeron86}. We revisit the absorber using preexisting data and new observations using $HST/$COS obtained under program ID 13398 as a part of our ``Multiphase Gaseous Halos" survey. The absorber shows a plethora of absorption lines arising from neutral and ionized metals and molecular hydrogen. Below we briefly summarize our main results:

\begin{enumerate}

\item The total \HI\ column density obtained by fitting the sub-DLA profile is $10^{19.50\pm0.15}$~\sqcm. The Lyman- and Werner-band absorption of \HH\ is detected up to the $J=3$ rotational level with a total $N(\HH) = 10^{16.36\pm0.08}$~\sqcm, corresponding to a molecular fraction of $\log f_{\hh} = -2.84\pm0.17$. The rotational excitation temperature of $T_{\rm ex}= 206\pm6$~K, obtained from the \HH\ level populations, reveals the presence of cold gas in the absorber. The measured $\log f_{\hh}$ and $T_{\rm ex}$ values are roughly consistent with the corresponding median values for the sample of low-$z$ \HH\ absorbers studied by \cite{Muzahid15a}.

\item Using a simple photoionization equilibrium model, we obtain a near-solar metallicity for the sub-DLA with oxygen abundance of $\rm [O/H] = -0.26\pm0.19$ and a density of $\log n_{\rm H} = -0.2$ (i.e., 0.6 particles cm$^{-3}$). Assuming the intrinsic $\rm [Fe/S]$ to be solar, we measured a relative dust-to-gas ratio of $\log \kappa = -0.45$ and a column density of dust in Fe of $N_{\rm Fe}^{\rm dust} \sim 10^{14.6}$~\sqcm.

\item Ionization-corrected gas-phase abundances of nine different elements, besides oxygen, are calculated. The abundances of S and Si are consistent with O, suggesting a lack of depletion in these elements. The observed $\rm [N/O]= -0.42\pm0.25$, along with $\rm[O/H]$, makes this system lie near the ``knee" of the $\rm[N/\alpha]$ versus $[\rm \alpha/H]$ plot \citep[see, e.g.,][]{Pettini08a}. The fact possibly implies that the synthesis of N in this absorber is contributed by both ``primary" and ``secondary" processes.

\item Both Fe and Mg show similar gas-phase abundances (i.e. $\rm [X/H]\sim-1.0$) with $\rm [Mg/Fe] = 0.13\pm0.23$ indicating a lack of (or no) significant $\alpha$-enhancement. Additionally, we observe Mn and Fe having identical abundances. Therefore, Fe, Mg, and Mn seem to show a moderate ($\gtrsim0.7$ dex) depletion. Ca, with $\rm [Ca/H] = -1.78\pm0.18$, is, however, heavily depleted.

\item The host galaxy of the sub-DLA is detected at an impact parameter of $\sim$48~kpc. The host galaxy has a $B$-band absolute magnitude of $M_{B} = -20.35$ (corresponding to $\sim$0.5$L_{\ast}$) and a $B-K$ color of 2.06 suggesting a moderate SFR \citep[]{Nielsen13a}. It has an inclination angle of $i\sim$48\degree\ on the plane of the sky. The QSO sightline is at an azimuthal angle of $\sim$15\degree\ from the projected major axis of the host galaxy.

\item Using models of \cite{Steidel02} we found that the \MgII\ absorption kinematics cannot be explained by gas co-rotating with an extended disk. The effective radius of the \HI\ disk of $\sim$13~kpc, derived from $M_{B}$ using the observed scaling relation of \cite{Lah09}, further refutes the origin of the absorber in an extended disk. Moreover, the high gas-phase metallicity of the absorber suggests that it is not tracing accreting materials. On the other hand, a large opening angle (2$\theta>$~150\degree) is required for an outflow from the central region of the host galaxy to intercept the QSO sightline at an azimuthal angle of $\sim$15\degree.

\item We favor a scenario in which the absorber originates in a satellite dwarf galaxy. A large $in~situ$ \HH\ formation timescale ($\sim$$5\times10^{9}$ yr) indicates that the molecular gas, instead, is stemming from stripped-off disk material of the satellite galaxy due to (i) ram pressure, as the  satellite moves into the dark matter potential of the bigger galaxy and$/$or (ii) tidal interaction between the satellite and the host galaxy. Interestingly, a dwarf galaxy is detected in the QSO PSF-subtracted $HST$ image of the field at an impact parameter of $\sim$12~kpc. Spectroscopic observations of the dwarf galaxy candidate are crucial for further comprehension on the origin of the sub-DLA.

\end{enumerate}

This exercise emphasizes the importance of detailed analysis of QSO absorbers on a case-by-case basis. While studies of large samples provide useful information on overall properties of the absorbers, they often tend to oversimplify the inherent intricacies of the problem, in particular, the origin(s) of the absorbers. We seek to perform similar analysis for all the low-$z$ \HH\ absorbers \citep[]{Muzahid15a} in the future with new observations using $HST$ and ground-based telescopes.

\vskip0.2cm 
S.M. thanks Dr. Hadi Rahmani for providing an independently reduced HIRES spectrum of the QSO. S.M. also thanks Rajeshwari Dutta and Dr. R. Srianand for useful suggestions. Support for this research was provided by NASA through grants $HST$ GO-13398 from the Space Telescope Science Institute, which is operated by the Association of Universities for Research in Astronomy, Inc., under NASA contract NAS5-26555. G.G.K. acknowledges the support of the Australian Research Council through the award of a Future Fellowship (FT140100933). Some of the data presented here were obtained at the W. M. Keck Observatory, which is operated as a scientific partnership among the California Institute of Technology, the University of California, and the National Aeronautics and Space Administration. The Observatory was made possible by the generous financial support of the W. M. Keck Foundation. Observations were supported by Swinburne Keck program 2015A\_W018E.

\vskip0.2cm 
\noindent 
{\it Facilities:}~$\it HST$(COS,~WFPC2),~Keck(HIRES, ESI)

\def\aj{AJ}%
\def\actaa{Acta Astron.}%
\def\araa{ARA\&A}%
\def\apj{ApJ}%
\def\apjl{ApJ}%
\def\apjs{ApJS}%
\def\ao{Appl.~Opt.}%
\def\apss{Ap\&SS}%
\def\aap{A\&A}%
\def\aapr{A\&A~Rev.}%
\def\aaps{A\&AS}%
\def\azh{AZh}%
\def\baas{BAAS}%
\def\bac{Bull. astr. Inst. Czechosl.}%
\def\caa{Chinese Astron. Astrophys.}%
\def\cjaa{Chinese J. Astron. Astrophys.}%
\def\icarus{Icarus}%
\def\jcap{J. Cosmology Astropart. Phys.}%
\def\jrasc{JRASC}%
\def\mnras{MNRAS}%
\def\memras{MmRAS}%
\def\na{New A}%
\def\nar{New A Rev.}%
\def\pasa{PASA}%
\def\pra{Phys.~Rev.~A}%
\def\prb{Phys.~Rev.~B}%
\def\prc{Phys.~Rev.~C}%
\def\prd{Phys.~Rev.~D}%
\def\pre{Phys.~Rev.~E}%
\def\prl{Phys.~Rev.~Lett.}%
\def\pasp{PASP}%
\def\pasj{PASJ}%
\def\qjras{QJRAS}%
\def\rmxaa{Rev. Mexicana Astron. Astrofis.}%
\def\skytel{S\&T}%
\def\solphys{Sol.~Phys.}%
\def\sovast{Soviet~Ast.}%
\def\ssr{Space~Sci.~Rev.}%
\def\zap{ZAp}%
\def\nat{Nature}%
\def\iaucirc{IAU~Circ.}%
\def\aplett{Astrophys.~Lett.}%
\def\apspr{Astrophys.~Space~Phys.~Res.}%
\def\bain{Bull.~Astron.~Inst.~Netherlands}%
\def\fcp{Fund.~Cosmic~Phys.}%
\def\gca{Geochim.~Cosmochim.~Acta}%
\def\grl{Geophys.~Res.~Lett.}%
\def\jcp{J.~Chem.~Phys.}%
\def\jgr{J.~Geophys.~Res.}%
\def\jqsrt{J.~Quant.~Spec.~Radiat.~Transf.}%
\def\memsai{Mem.~Soc.~Astron.~Italiana}%
\def\nphysa{Nucl.~Phys.~A}%
\def\physrep{Phys.~Rep.}%
\def\physscr{Phys.~Scr}%
\def\planss{Planet.~Space~Sci.}%
\def\procspie{Proc.~SPIE}%
\let\astap=\aap
\let\apjlett=\apjl
\let\apjsupp=\apjs
\let\applopt=\ao
\bibliographystyle{apj}
\bibliography{/home/sowgat/Work/softwares/LaTeX/mybib}

\begin{thebibliography}{}
\expandafter\ifx\csname natexlab\endcsname\relax\def\natexlab#1{#1}\fi

\bibitem[{{Akerman} {et~al.}(2004){Akerman}, {Carigi}, {Nissen}, {Pettini}, \&
  {Asplund}}]{Akerman04}
{Akerman}, C.~J., {Carigi}, L., {Nissen}, P.~E., {Pettini}, M., \& {Asplund},
  M. 2004, \aap, 414, 931

\bibitem[{{Alatalo} {et~al.}(2011){Alatalo}, {Blitz}, {Young}, {Davis},
  {Bureau}, {Lopez}, {Cappellari}, {Scott}, {Shapiro}, {Crocker},
  {Mart{\'{\i}}n}, {Bois}, {Bournaud}, {Davies}, {de Zeeuw}, {Duc}, {Emsellem},
  {Falc{\'o}n-Barroso}, {Khochfar}, {Krajnovi{\'c}}, {Kuntschner}, {Lablanche},
  {McDermid}, {Morganti}, {Naab}, {Oosterloo}, {Sarzi}, {Serra}, \&
  {Weijmans}}]{Alatalo11}
{Alatalo}, K., {Blitz}, L., {Young}, L.~M., {et~al.} 2011, \apj, 735, 88

\bibitem[{{Asplund} {et~al.}(2009){Asplund}, {Grevesse}, {Sauval}, \&
  {Scott}}]{Asplund09}
{Asplund}, M., {Grevesse}, N., {Sauval}, A.~J., \& {Scott}, P. 2009, \araa, 47,
  481

\bibitem[{{Bergeron}(1986)}]{Bergeron86}
{Bergeron}, J. 1986, \aap, 155, L8

\bibitem[{{Bertin} \& {Arnouts}(1996)}]{Bertin96}
{Bertin}, E., \& {Arnouts}, S. 1996, \aaps, 117, 393

\bibitem[{{Black} {et~al.}(1987){Black}, {Chaffee}, \& {Foltz}}]{Black87}
{Black}, J.~H., {Chaffee}, F.~H., \& {Foltz}, C.~B. 1987, \apj, 317, 442

\bibitem[{{Bordoloi} {et~al.}(2014){Bordoloi}, {Lilly}, {Kacprzak}, \&
  {Churchill}}]{Bordoloi14c}
{Bordoloi}, R., {Lilly}, S.~J., {Kacprzak}, G.~G., \& {Churchill}, C.~W. 2014,
  \apj, 784, 108

\bibitem[{{Bouwens} {et~al.}(2011){Bouwens}, {Illingworth}, {Oesch},
  {Labb{\'e}}, {Trenti}, {van Dokkum}, {Franx}, {Stiavelli}, {Carollo},
  {Magee}, \& {Gonzalez}}]{Bouwens11}
{Bouwens}, R.~J., {Illingworth}, G.~D., {Oesch}, P.~A., {et~al.} 2011, \apj,
  737, 90

\bibitem[{{Chen} {et~al.}(2005){Chen}, {Kennicutt}, \& {Rauch}}]{Chen05}
{Chen}, H.-W., {Kennicutt}, Jr., R.~C., \& {Rauch}, M. 2005, \apj, 620, 703

\bibitem[{{Churchill} {et~al.}(2000){Churchill}, {Mellon}, {Charlton},
  {Jannuzi}, {Kirhakos}, {Steidel}, \& {Schneider}}]{Churchill00a}
{Churchill}, C.~W., {Mellon}, R.~R., {Charlton}, J.~C., {et~al.} 2000, \apjs,
  130, 91

\bibitem[{{Crighton} {et~al.}(2013){Crighton}, {Bechtold}, {Carswell},
  {Dav{\'e}}, {Foltz}, {Jannuzi}, {Morris}, {O'Meara}, {Prochaska}, {Schaye},
  \& {Tejos}}]{Crighton13}
{Crighton}, N.~H.~M., {Bechtold}, J., {Carswell}, R.~F., {et~al.} 2013, \mnras,
  433, 178

\bibitem[{{Danforth} {et~al.}(2010){Danforth}, {Stocke}, \&
  {Shull}}]{Danforth10}
{Danforth}, C.~W., {Stocke}, J.~T., \& {Shull}, J.~M. 2010, \apj, 710, 613

\bibitem[{{D'Onghia} \& {Fox}(2015)}]{DOnghia15}
{D'Onghia}, E., \& {Fox}, A.~J. 2015, ArXiv e-prints, arXiv:1511.05853

\bibitem[{{Dutta} {et~al.}(2015){Dutta}, {Srianand}, {Muzahid}, {Gupta},
  {Momjian}, \& {Charlton}}]{Dutta15}
{Dutta}, R., {Srianand}, R., {Muzahid}, S., {et~al.} 2015, \mnras, 448, 3718

\bibitem[{{Ferland} {et~al.}(2013){Ferland}, {Porter}, {van Hoof}, {Williams},
  {Abel}, {Lykins}, {Shaw}, {Henney}, \& {Stancil}}]{Ferland13}
{Ferland}, G.~J., {Porter}, R.~L., {van Hoof}, P.~A.~M., {et~al.} 2013, \rmxaa,
  49, 137

\bibitem[{{Geach} {et~al.}(2014){Geach}, {Hickox}, {Diamond-Stanic}, {Krips},
  {Rudnick}, {Tremonti}, {Sell}, {Coil}, \& {Moustakas}}]{Geach14Nat}
{Geach}, J.~E., {Hickox}, R.~C., {Diamond-Stanic}, A.~M., {et~al.} 2014, \nat,
  516, 68

\bibitem[{{Gillmon} {et~al.}(2006){Gillmon}, {Shull}, {Tumlinson}, \&
  {Danforth}}]{Gillmon06}
{Gillmon}, K., {Shull}, J.~M., {Tumlinson}, J., \& {Danforth}, C. 2006, \apj,
  636, 891

\bibitem[{{Gnat} \& {Sternberg}(2007)}]{Gnat07}
{Gnat}, O., \& {Sternberg}, A. 2007, \apjs, 168, 213

\bibitem[{{Green} {et~al.}(2012){Green}, {Froning}, {Osterman}, {Ebbets},
  {Heap}, {Leitherer}, {Linsky}, {Savage}, {Sembach}, {Shull}, {Siegmund},
  {Snow}, {Spencer}, {Stern}, {Stocke}, {Welsh}, {B{\'e}land}, {Burgh},
  {Danforth}, {France}, {Keeney}, {McPhate}, {Penton}, {Andrews},
  {Brownsberger}, {Morse}, \& {Wilkinson}}]{Green12}
{Green}, J.~C., {Froning}, C.~S., {Osterman}, S., {et~al.} 2012, \apj, 744, 60

\bibitem[{{Gringel} {et~al.}(2000){Gringel}, {Barnstedt}, {de Boer}, {Grewing},
  {Kappelmann}, \& {Richter}}]{Gringel00}
{Gringel}, W., {Barnstedt}, J., {de Boer}, K.~S., {et~al.} 2000, \aap, 358, L37

\bibitem[{{Haardt} \& {Madau}(2012)}]{Haardt12}
{Haardt}, F., \& {Madau}, P. 2012, \apj, 746, 125

\bibitem[{{Hamann} \& {Ferland}(1993)}]{Hamann93}
{Hamann}, F., \& {Ferland}, G. 1993, \apj, 418, 11

\bibitem[{{Hirashita} {et~al.}(2003){Hirashita}, {Ferrara}, {Wada}, \&
  {Richter}}]{Hirashita03}
{Hirashita}, H., {Ferrara}, A., {Wada}, K., \& {Richter}, P. 2003, \mnras, 341,
  L18

\bibitem[{{Hjelm} \& {Lindblad}(1996)}]{Hjelm96}
{Hjelm}, M., \& {Lindblad}, P.~O. 1996, \aap, 305, 727

\bibitem[{{Hollenbach} \& {Salpeter}(1971)}]{Hollenbach71a}
{Hollenbach}, D., \& {Salpeter}, E.~E. 1971, \apj, 163, 155

\bibitem[{{Jura}(1974)}]{Jura74b}
{Jura}, M. 1974, \apj, 191, 375

\bibitem[{{Kacprzak} {et~al.}(2010){Kacprzak}, {Churchill}, {Ceverino},
  {Steidel}, {Klypin}, \& {Murphy}}]{Kacprzak10a}
{Kacprzak}, G.~G., {Churchill}, C.~W., {Ceverino}, D., {et~al.} 2010, \apj,
  711, 533

\bibitem[{{Kacprzak} {et~al.}(2011){Kacprzak}, {Churchill}, {Evans}, {Murphy},
  \& {Steidel}}]{Kacprzak11}
{Kacprzak}, G.~G., {Churchill}, C.~W., {Evans}, J.~L., {Murphy}, M.~T., \&
  {Steidel}, C.~C. 2011, \mnras, 416, 3118

\bibitem[{{Kacprzak} {et~al.}(2012{\natexlab{a}}){Kacprzak}, {Churchill}, \&
  {Nielsen}}]{Kacprzak12a}
{Kacprzak}, G.~G., {Churchill}, C.~W., \& {Nielsen}, N.~M. 2012{\natexlab{a}},
  \apjl, 760, L7

\bibitem[{{Kacprzak} {et~al.}(2012{\natexlab{b}}){Kacprzak}, {Churchill},
  {Steidel}, {Spitler}, \& {Holtzman}}]{Kacprzak12b}
{Kacprzak}, G.~G., {Churchill}, C.~W., {Steidel}, C.~C., {Spitler}, L.~R., \&
  {Holtzman}, J.~A. 2012{\natexlab{b}}, \mnras, 427, 3029

\bibitem[{{Kacprzak} {et~al.}(2015){Kacprzak}, {Muzahid}, {Churchill},
  {Nielsen}, \& {Charlton}}]{Kacprzak15}
{Kacprzak}, G.~G., {Muzahid}, S., {Churchill}, C.~W., {Nielsen}, N.~M., \&
  {Charlton}, J.~C. 2015, \apj, 815, 22

\bibitem[{{Kanekar} \& {Chengalur}(2003)}]{Kanekar03}
{Kanekar}, N., \& {Chengalur}, J.~N. 2003, \aap, 399, 857

\bibitem[{{Kriss}(2011)}]{Kriss11}
{Kriss}, G.~A. 2011, {Improved Medium Resolution Line Spread Functions for COS
  FUV Spectra}, Tech. rep.

\bibitem[{{Lah} {et~al.}(2009){Lah}, {Pracy}, {Chengalur}, {Briggs}, {Colless},
  {de Propris}, {Ferris}, {Schmidt}, \& {Tucker}}]{Lah09}
{Lah}, P., {Pracy}, M.~B., {Chengalur}, J.~N., {et~al.} 2009, \mnras, 399, 1447

\bibitem[{{Ledoux} {et~al.}(2003){Ledoux}, {Petitjean}, \&
  {Srianand}}]{Ledoux03}
{Ledoux}, C., {Petitjean}, P., \& {Srianand}, R. 2003, \mnras, 346, 209

\bibitem[{{Lehner}(2002)}]{Lehner02}
{Lehner}, N. 2002, \apj, 578, 126

\bibitem[{{Lehner} {et~al.}(2013){Lehner}, {Howk}, {Tripp}, {Tumlinson},
  {Prochaska}, {O'Meara}, {Thom}, {Werk}, {Fox}, \& {Ribaudo}}]{Lehner13}
{Lehner}, N., {Howk}, J.~C., {Tripp}, T.~M., {et~al.} 2013, \apj, 770, 138

\bibitem[{{M{\"u}ller-S{\'a}nchez} {et~al.}(2011){M{\"u}ller-S{\'a}nchez},
  {Prieto}, {Hicks}, {Vives-Arias}, {Davies}, {Malkan}, {Tacconi}, \&
  {Genzel}}]{Mullar11}
{M{\"u}ller-S{\'a}nchez}, F., {Prieto}, M.~A., {Hicks}, E.~K.~S., {et~al.}
  2011, \apj, 739, 69

\bibitem[{{Muzahid}(2014)}]{Muzahid14}
{Muzahid}, S. 2014, \apj, 784, 5

\bibitem[{{Muzahid} {et~al.}(2015{\natexlab{a}}){Muzahid}, {Kacprzak},
  {Churchill}, {Charlton}, {Nielsen}, {Mathes}, \&
  {Trujillo-Gomez}}]{Muzahid15b}
{Muzahid}, S., {Kacprzak}, G.~G., {Churchill}, C.~W., {et~al.}
  2015{\natexlab{a}}, \apj, 811, 132

\bibitem[{{Muzahid} {et~al.}(2015{\natexlab{b}}){Muzahid}, {Srianand}, \&
  {Charlton}}]{Muzahid15a}
{Muzahid}, S., {Srianand}, R., \& {Charlton}, J. 2015{\natexlab{b}}, \mnras,
  448, 2840

\bibitem[{{Narayanan} {et~al.}(2010){Narayanan}, {Savage}, \&
  {Wakker}}]{Narayanan10}
{Narayanan}, A., {Savage}, B.~D., \& {Wakker}, B.~P. 2010, \apj, 712, 1443

\bibitem[{{Nielsen} {et~al.}(2013){Nielsen}, {Churchill}, {Kacprzak}, \&
  {Murphy}}]{Nielsen13a}
{Nielsen}, N.~M., {Churchill}, C.~W., {Kacprzak}, G.~G., \& {Murphy}, M.~T.
  2013, \apj, 776, 114

\bibitem[{{Noterdaeme} {et~al.}(2008){Noterdaeme}, {Ledoux}, {Petitjean}, \&
  {Srianand}}]{Noterdaeme08}
{Noterdaeme}, P., {Ledoux}, C., {Petitjean}, P., \& {Srianand}, R. 2008, \aap,
  481, 327

\bibitem[{{Oliveira} {et~al.}(2014){Oliveira}, {Sembach}, {Tumlinson},
  {O'Meara}, \& {Thom}}]{Oliveira14}
{Oliveira}, C.~M., {Sembach}, K.~R., {Tumlinson}, J., {O'Meara}, J., \& {Thom},
  C. 2014, \apj, 783, 22

\bibitem[{{Osterman} {et~al.}(2011){Osterman}, {Green}, {Froning},
  {B{\'e}land}, {Burgh}, {France}, {Penton}, {Delker}, {Ebbets}, {Sahnow},
  {Bacinski}, {Kimble}, {Andrews}, {Wilkinson}, {McPhate}, {Siegmund}, {Ake},
  {Aloisi}, {Biagetti}, {Diaz}, {Dixon}, {Friedman}, {Ghavamian}, {Goudfrooij},
  {Hartig}, {Keyes}, {Lennon}, {Massa}, {Niemi}, {Oliveira}, {Osten},
  {Proffitt}, {Smith}, \& {Soderblom}}]{Osterman11}
{Osterman}, S., {Green}, J., {Froning}, C., {et~al.} 2011, \apss, 335, 257

\bibitem[{{Petitjean} {et~al.}(2006){Petitjean}, {Ledoux}, {Noterdaeme}, \&
  {Srianand}}]{Petitjean06}
{Petitjean}, P., {Ledoux}, C., {Noterdaeme}, P., \& {Srianand}, R. 2006, \aap,
  456, L9

\bibitem[{{Petitjean} {et~al.}(1996){Petitjean}, {Riediger}, \&
  {Rauch}}]{Petitjean96}
{Petitjean}, P., {Riediger}, R., \& {Rauch}, M. 1996, \aap, 307, 417

\bibitem[{{Pettini} \& {Pagel}(2004)}]{Pettini04}
{Pettini}, M., \& {Pagel}, B.~E.~J. 2004, \mnras, 348, L59

\bibitem[{{Pettini} {et~al.}(2008){Pettini}, {Zych}, {Steidel}, \&
  {Chaffee}}]{Pettini08a}
{Pettini}, M., {Zych}, B.~J., {Steidel}, C.~C., \& {Chaffee}, F.~H. 2008,
  \mnras, 385, 2011

\bibitem[{{Rafelski} {et~al.}(2012){Rafelski}, {Wolfe}, {Prochaska},
  {Neeleman}, \& {Mendez}}]{Rafelski12}
{Rafelski}, M., {Wolfe}, A.~M., {Prochaska}, J.~X., {Neeleman}, M., \&
  {Mendez}, A.~J. 2012, \apj, 755, 89

\bibitem[{{Ribaudo} {et~al.}(2011){Ribaudo}, {Lehner}, {Howk}, {Werk}, {Tripp},
  {Prochaska}, {Meiring}, \& {Tumlinson}}]{Ribaudo11}
{Ribaudo}, J., {Lehner}, N., {Howk}, J.~C., {et~al.} 2011, \apj, 743, 207

\bibitem[{{Richter} {et~al.}(1999){Richter}, {de Boer}, {Widmann},
  {Kappelmann}, {Gringel}, {Grewing}, \& {Barnstedt}}]{Richter99b}
{Richter}, P., {de Boer}, K.~S., {Widmann}, H., {et~al.} 1999, \nat, 402, 386

\bibitem[{{Richter} {et~al.}(2001){Richter}, {Sembach}, {Wakker}, \&
  {Savage}}]{Richter01a}
{Richter}, P., {Sembach}, K.~R., {Wakker}, B.~P., \& {Savage}, B.~D. 2001,
  \apjl, 562, L181

\bibitem[{{Richter} {et~al.}(2003){Richter}, {Wakker}, {Savage}, \&
  {Sembach}}]{Richter03}
{Richter}, P., {Wakker}, B.~P., {Savage}, B.~D., \& {Sembach}, K.~R. 2003,
  \apj, 586, 230

\bibitem[{{Rigby} {et~al.}(2002){Rigby}, {Charlton}, \& {Churchill}}]{Rigby02}
{Rigby}, J.~R., {Charlton}, J.~C., \& {Churchill}, C.~W. 2002, \apj, 565, 743

\bibitem[{{Roy} {et~al.}(2006){Roy}, {Chengalur}, \& {Srianand}}]{Roy06}
{Roy}, N., {Chengalur}, J.~N., \& {Srianand}, R. 2006, \mnras, 365, L1

\bibitem[{{Savage} {et~al.}(1977){Savage}, {Bohlin}, {Drake}, \&
  {Budich}}]{Savage77}
{Savage}, B.~D., {Bohlin}, R.~C., {Drake}, J.~F., \& {Budich}, W. 1977, \apj,
  216, 291

\bibitem[{{Savage} {et~al.}(2011){Savage}, {Lehner}, \&
  {Narayanan}}]{Savage11a}
{Savage}, B.~D., {Lehner}, N., \& {Narayanan}, A. 2011, \apj, 743, 180

\bibitem[{{Savage} \& {Sembach}(1991)}]{Savage91}
{Savage}, B.~D., \& {Sembach}, K.~R. 1991, \apj, 379, 245

\bibitem[{{Savage} \& {Sembach}(1996)}]{Savage96a}
---. 1996, \araa, 34, 279

\bibitem[{{Sembach} {et~al.}(2001){Sembach}, {Howk}, {Savage}, \&
  {Shull}}]{Sembach01}
{Sembach}, K.~R., {Howk}, J.~C., {Savage}, B.~D., \& {Shull}, J.~M. 2001, \aj,
  121, 992

\bibitem[{{Shaw} {et~al.}(2005){Shaw}, {Ferland}, {Abel}, {Stancil}, \& {van
  Hoof}}]{Shaw05}
{Shaw}, G., {Ferland}, G.~J., {Abel}, N.~P., {Stancil}, P.~C., \& {van Hoof},
  P.~A.~M. 2005, \apj, 624, 794

\bibitem[{{Sheinis} {et~al.}(2002){Sheinis}, {Bolte}, {Epps}, {Kibrick},
  {Miller}, {Radovan}, {Bigelow}, \& {Sutin}}]{Sheinis02}
{Sheinis}, A.~I., {Bolte}, M., {Epps}, H.~W., {et~al.} 2002, \pasp, 114, 851

\bibitem[{{Shull} \& {Beckwith}(1982)}]{Shull82}
{Shull}, J.~M., \& {Beckwith}, S. 1982, \araa, 20, 163

\bibitem[{{Simard} {et~al.}(2002){Simard}, {Willmer}, {Vogt}, {Sarajedini},
  {Phillips}, {Weiner}, {Koo}, {Im}, {Illingworth}, \& {Faber}}]{Simard02}
{Simard}, L., {Willmer}, C.~N.~A., {Vogt}, N.~P., {et~al.} 2002, \apjs, 142, 1

\bibitem[{{Snow} \& {McCall}(2006)}]{Snow06}
{Snow}, T.~P., \& {McCall}, B.~J. 2006, \araa, 44, 367

\bibitem[{{Som} {et~al.}(2013){Som}, {Kulkarni}, {Meiring}, {York},
  {P{\'e}roux}, {Khare}, \& {Lauroesch}}]{Som13}
{Som}, D., {Kulkarni}, V.~P., {Meiring}, J., {et~al.} 2013, \mnras, 435, 1469

\bibitem[{{Spitzer} \& {Jenkins}(1975)}]{Spitzer75}
{Spitzer}, Jr., L., \& {Jenkins}, E.~B. 1975, \araa, 13, 133

\bibitem[{{Srianand} {et~al.}(2012){Srianand}, {Gupta}, {Petitjean},
  {Noterdaeme}, {Ledoux}, {Salter}, \& {Saikia}}]{Srianand12}
{Srianand}, R., {Gupta}, N., {Petitjean}, P., {et~al.} 2012, \mnras, 421, 651

\bibitem[{{Srianand} {et~al.}(2008){Srianand}, {Noterdaeme}, {Ledoux}, \&
  {Petitjean}}]{Srianand08}
{Srianand}, R., {Noterdaeme}, P., {Ledoux}, C., \& {Petitjean}, P. 2008, \aap,
  482, L39

\bibitem[{{Srianand} {et~al.}(2005){Srianand}, {Petitjean}, {Ledoux},
  {Ferland}, \& {Shaw}}]{Srianand05}
{Srianand}, R., {Petitjean}, P., {Ledoux}, C., {Ferland}, G., \& {Shaw}, G.
  2005, \mnras, 362, 549

\bibitem[{{Srianand} {et~al.}(2014){Srianand}, {Rahmani}, {Muzahid}, \&
  {Mohan}}]{Srianand14}
{Srianand}, R., {Rahmani}, H., {Muzahid}, S., \& {Mohan}, V. 2014, \mnras, 443,
  3318

\bibitem[{{Steidel} {et~al.}(2002){Steidel}, {Kollmeier}, {Shapley},
  {Churchill}, {Dickinson}, \& {Pettini}}]{Steidel02}
{Steidel}, C.~C., {Kollmeier}, J.~A., {Shapley}, A.~E., {et~al.} 2002, \apj,
  570, 526

\bibitem[{{Tripp} {et~al.}(2011){Tripp}, {Meiring}, {Prochaska}, {Willmer},
  {Howk}, {Werk}, {Jenkins}, {Bowen}, {Lehner}, {Sembach}, {Thom}, \&
  {Tumlinson}}]{Tripp11}
{Tripp}, T.~M., {Meiring}, J.~D., {Prochaska}, J.~X., {et~al.} 2011, Science,
  334, 952

\bibitem[{{Tumlinson} {et~al.}(2002){Tumlinson}, {Shull}, {Rachford},
  {Browning}, {Snow}, {Fullerton}, {Jenkins}, {Savage}, {Crowther}, {Moos},
  {Sembach}, {Sonneborn}, \& {York}}]{Tumlinson02}
{Tumlinson}, J., {Shull}, J.~M., {Rachford}, B.~L., {et~al.} 2002, \apj, 566,
  857

\bibitem[{{Veilleux} {et~al.}(2005){Veilleux}, {Cecil}, \&
  {Bland-Hawthorn}}]{Veilleux05}
{Veilleux}, S., {Cecil}, G., \& {Bland-Hawthorn}, J. 2005, \araa, 43, 769

\bibitem[{{Veilleux} {et~al.}(2001){Veilleux}, {Shopbell}, \&
  {Miller}}]{Veilleux01}
{Veilleux}, S., {Shopbell}, P.~L., \& {Miller}, S.~T. 2001, \aj, 121, 198

\bibitem[{{Wakker}(2006)}]{Wakker06}
{Wakker}, B.~P. 2006, \apjs, 163, 282

\bibitem[{{Welty} {et~al.}(1999){Welty}, {Hobbs}, {Lauroesch}, {Morton},
  {Spitzer}, \& {York}}]{Welty99a}
{Welty}, D.~E., {Hobbs}, L.~M., {Lauroesch}, J.~T., {et~al.} 1999, \apjs, 124,
  465

\bibitem[{{Welty} {et~al.}(2012){Welty}, {Xue}, \& {Wong}}]{Welty12}
{Welty}, D.~E., {Xue}, R., \& {Wong}, T. 2012, \apj, 745, 173

\bibitem[{{Werk} {et~al.}(2014){Werk}, {Prochaska}, {Tumlinson}, {Peeples},
  {Tripp}, {Fox}, {Lehner}, {Thom}, {O'Meara}, {Ford}, {Bordoloi}, {Katz},
  {Tejos}, {Oppenheimer}, {Dav{\'e}}, \& {Weinberg}}]{Werk14}
{Werk}, J.~K., {Prochaska}, J.~X., {Tumlinson}, J., {et~al.} 2014, \apj, 792, 8

\bibitem[{{Wiersma} {et~al.}(2009){Wiersma}, {Schaye}, \& {Smith}}]{Wiersma09}
{Wiersma}, R.~P.~C., {Schaye}, J., \& {Smith}, B.~D. 2009, \mnras, 393, 99

\bibitem[{{Williams} {et~al.}(1998){Williams}, {Bergin}, {Caselli}, {Myers}, \&
  {Plume}}]{Williams98}
{Williams}, J.~P., {Bergin}, E.~A., {Caselli}, P., {Myers}, P.~C., \& {Plume},
  R. 1998, \apj, 503, 689

\bibitem[{{Wolfe} {et~al.}(2003){Wolfe}, {Prochaska}, \& {Gawiser}}]{Wolfe03}
{Wolfe}, A.~M., {Prochaska}, J.~X., \& {Gawiser}, E. 2003, \apj, 593, 215

\end{thebibliography}

\appendix

\section{Analysis of the \OVI\ absorption}    
\label{sec:highion}        

\begin{figure}[h] 
\centerline{
\vbox{
\centerline{\hbox{ 
\includegraphics[width=0.48\textwidth,angle=00]{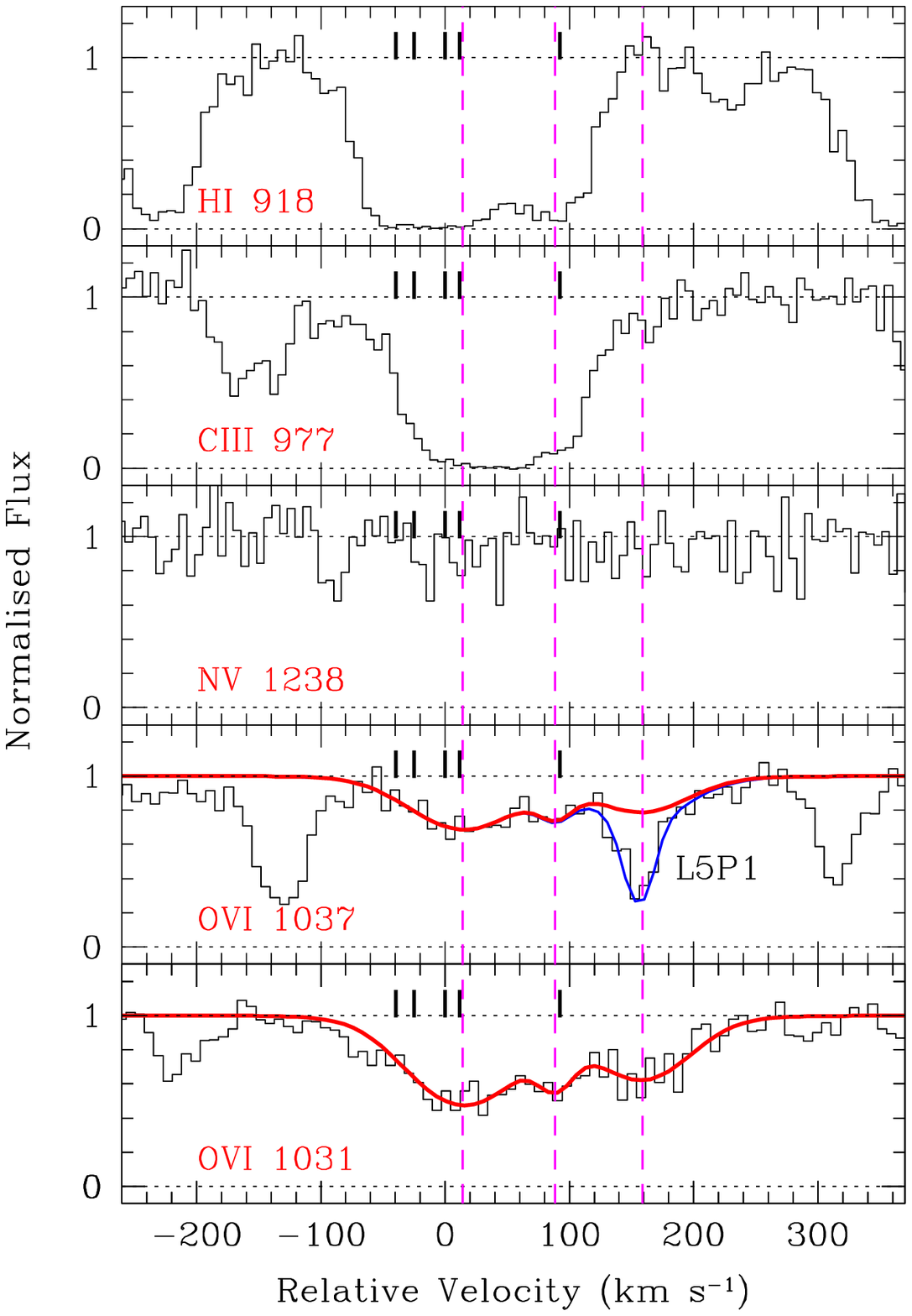}  
}} 
}}  
\caption{Voigt profile decomposition for the \OVI\ absorption. The absorption profiles are plotted in velocity with respect to \zabs~$=$~0.429805. The smooth red curves are the best-fitting Voigt profiles over-plotted on top of data (black histogram). The line centroids of \OVI\ components are marked by vertical dotted lines. The solid black ticks represent the centroids of \MgII\ components. The \OVI~$\lambda$1037 profile is blended with one of the \HH\ transitions from the $J=1$ level. \HI~$\lambda918$ and \CIII\ absorption profiles are shown just for references.}    
\label{fig:highion} 
\end{figure}

In this appendix we discuss the properties of the high-ionization gas-phase traced by the \OVI\ doublet. The Voigt profile decomposition of the detected \OVI\ is shown in Figure~\ref{fig:highion}. A minimum of three components are required for fitting both the doublet members adequately. Note that the \OVI~$\lambda$1037 is blended with the $\rm L5P1$ transition of \HH. The doublets are fitted simultaneously, taking into account the contribution of the \HH\ line. The fit parameters are summarized in Table~\ref{tab:highion}. As mentioned previously, \CIV\ and \NV\ are not detected in this system. Formal 3$\sigma$ upper limits on column densities for these ions are also presented in the table.     

\begin{figure}[h] 
\centerline{
\vbox{
\centerline{\hbox{ 
\includegraphics[width=0.55\textwidth,angle=00]{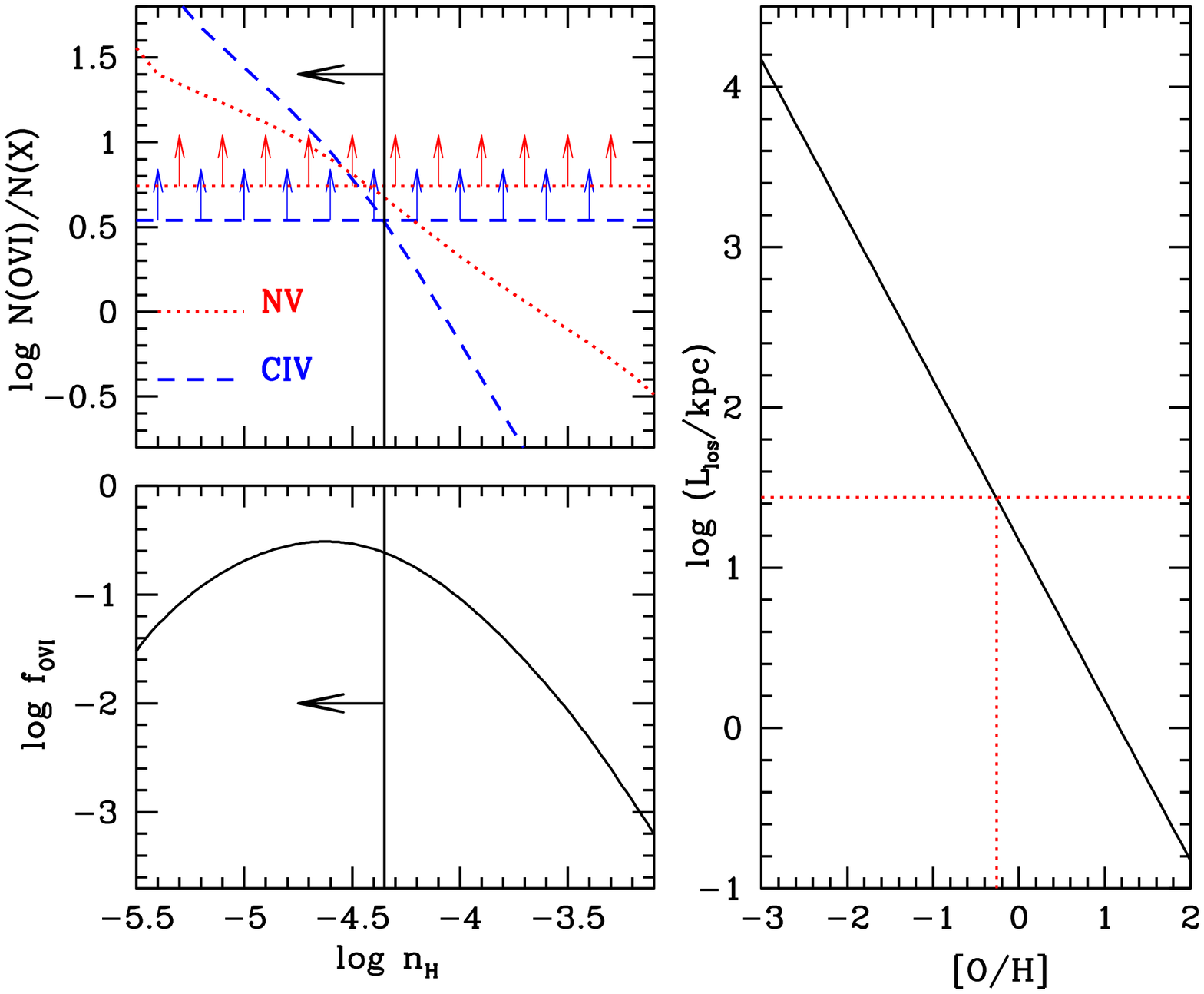}  
}} 
}}  
\caption{Photoionization model for the \OVI. Top left: column density ratios against the absorber's density. The vertical line followed by an arrow indicates the allowed density range. Bottom left: ionization fraction, $f_{\OVI}$, against the density. The $f_{\OVI}$ gradually increases with decreasing density to its peak at $\log n_{\rm H}\simeq-4.5$ and then decreases again. Right: variation of line-of-sight thickness with absorber's metallicity for a gas cloud with $\log N(\OVI)=14.34$, $\log n_{\rm H}=-4.5$ and $\log f_{\OVI} = -0.50$ (the peak value). The dotted horizontal line represents $L_{\rm los}$ of 28~kpc corresponding to $\rm [O/H]=-0.26$.}       
\label{fig:modhighion} 
\end{figure}

We run constant-density {\sc cloudy} simulations under optically thin conditions with a stopping $N(\HI)$ of $10^{14}$~\sqcm\ and the extragalactic UV background radiation at $z=0.42$ \citep[]{Haardt12} as the ionizing continuum. In Figure~\ref{fig:modhighion} we show the results of our photoionization model. In the top left corner of the figure the variations of $N(\OVI)/N(\CIV)$ and $N(\OVI)/N(\NV)$ ratios with density are shown. The observed lower limits on these ratios suggest $\log n_{\rm H}<-4.4$, provided that the relative abundances of N, C, and O are solar. Underabundance of C or N with respect to O would require even lower densities. From the bottom left panel it is evident that the \OVI\ ionization fraction, $f_{\OVI}$, peaks at $\log n_{\rm H}\simeq-4.5$ and decreases for $\log n_{\rm H}>-5.0$. The \OVI\ column density can be expressed as $N(\OVI) = f_{\OVI}~{\rm 10^{[O/H]}~(O/H)_{\odot}}~n_{\rm H}~L_{\rm los}$, where $L_{\rm los}$ is the line-of-sight thickness of the absorbing gas. Using the maximum $f_{\OVI}$ value, i.e. at $n_{\rm H} = -4.5$, it is now possible to examine how $L_{\rm los}$ changes with metallicity, $\rm [O/H]$, of the absorber for a given $N(\OVI)$. The corresponding plot, for the strongest \OVI\ component with $\log N(\OVI)=14.34$, is shown in the right panel of Figure~\ref{fig:modhighion}. It is clear that a metallicity much lower than $1/10$ of solar (i.e. $\rm [O/H] = -1.0$) is not permitted as it requires unreasonably large ($>$Mpc) line-of-sight thickness. If the high-ionization gas-phase has a metallicity similar to the \HH-bearing gas-phase (i.e. $\rm [O/H]=-0.26$) then the deduced $L_{\rm los}$ is 28~kpc. Lower $f_{\OVI}$ values at lower densities, since higher densities are forbidden, would require slightly higher values of $L_{\rm los}$. But the thickness could be smaller for higher metallicities. Therefore, the detected \OVI\ can have a reasonable solution under photoionization equilibrium provided that $\rm [O/H] > -1.0$. The only caveat here is that we assumed that the $f_{\OVI}$ is independent of metallicity, which is not strictly true for very high metallicities (e.g. ${\rm [O/H]}>0.5$). Nonetheless, the effect is small, and the cloud size becomes unreasonably large only at lower metallicities.   

Finally, we note that if the \OVI\ is in collisional ionization equilibrium \citep[]{Gnat07}, then the lower limits on the $N(\OVI)/N(\CIV)$ and $N(\OVI)/N(\NV)$ ratios require a gas temperature of $\log T > 5.4$, which is consistent with the maximum allowed temperatures evaluated from $b(\OVI)$ in Table~\ref{tab:highion}. Thus, the high-ions can arise either from photoionization or from collisional ionization.

\begin{table}[h] 
\begin{center}  
\caption{Voigt profile fit parameters for the \OVI\ absorption.}         
\begin{tabular}{ccccc}    
\hline 
Ion    &           \zabs\          &   $b$ (\kms)  &  $\log~(N/$\sqcm) &  $\log~(T_{\rm max}/K)^{1}$  \\            
\hline 
\OVI  &   0.429872$\pm$0.000018  &  53$\pm$7  &  14.34$\pm$0.04  &  6.54  \\  
\NV   &                          &            & $<$13.6          &        \\ 
\CIV  &                          &            & $<$13.8          &        \\ \\ 
\OVI  &   0.430227$\pm$0.000019  &  15$\pm$8  &  13.65$\pm$0.17  &  5.71  \\ 
\NV   &                          &            & $<$13.4          &        \\ 
\CIV  &                          &            & $<$13.5          &        \\ \\   
\OVI  &   0.430561$\pm$0.000026  &  47$\pm$9  &  14.08$\pm$0.06  &  6.48  \\ 
\NV   &                          &            & $<$13.5          &        \\ 
\CIV  &                          &            & $<$13.6          &        \\      
\hline   
\label{tab:highion}      
\end{tabular}
~\\ 
Note-- $^{1}$Maximum logarithmic gas temperature calculated from $b(\OVI)$ and the corresponding error.   
\end{center} 
\end{table}   

\end{document}